\documentclass{article}
\usepackage{algorithm}
\usepackage{algpseudocode}
\usepackage{authblk}
\usepackage{graphicx} 
\usepackage{amsmath}
\usepackage{appendix}
\usepackage{xcolor}

\usepackage{subcaption}
\usepackage{url}

\title{Towards neural reinforcement learning for large deviations in nonequilibrium systems with memory}

\author{Venkata D. Pamulaparthy\footnote{venkata.pamulaparthy.22@ucl.ac.uk } }

\author{Rosemary J. Harris\footnote{rosemary.j.harris@ucl.ac.uk}}
\affil{Department of Mathematics, University College London, Gower Street,
London
WC1E 6BT, United Kingdom}

\begin{document}

\maketitle

\begin{abstract}
We introduce a reinforcement learning method for a class of non-Markov systems; our approach extends the actor-critic framework given by Rose et al.\ [New J.\ Phys.\ \textbf{23}\ 013013 (2021)] for obtaining scaled cumulant generating functions characterizing the fluctuations. The actor-critic is implemented using  neural networks; a
 particular innovation in our method is the use of an additional neural policy
 for processing memory variables. We demonstrate results for current fluctuations in
 various memory-dependent models with special focus on semi-Markov systems
 where the dynamics is controlled by nonexponential interevent waiting time
 distributions.

\end{abstract}

\tableofcontents

\section{Introduction}

Large-deviation theory provides a foundational framework for stochastic systems out of equilibrium \cite{Derrida_2007, touchette1,touchette2,Ellis1,Ellis2006}. The framework defines key quantities analogous to the free energy and the entropy in equilibrium statistical physics, the \textit{scaled cumulant generating function} (SCGF) and the \textit{rate function}; these characterize fluctuations of time-averaged observables and are therefore relevant for understanding rare events (atypical behavior) in applications ranging from biology to finance \cite{large_deviations_bio,large_deviations_finance}.
For systems without memory (Markov processes), there is a concrete theoretical procedure based on spectral calculations to obtain the large-deviation quantities. However, except in simple toy models, this analytical procedure becomes complicated, necessitating numerical methods.

In recent years, many computational techniques have been developed to simulate and analyze rare events in systems out of equilibrium. Prominent approaches include cloning \cite{Lecomte_2007, giardina} which is based on population dynamics of trajectories, and transition path sampling which relies on importance sampling of trajectories \cite{Bolhuis,Asghar2024}. Notably, there has still been limited exploration of large-deviation computations for non-Markov systems, even though memory plays an important role in many real-world scenarios \cite{long_range,inter_cellular_memory,infection_non_Markov, glass_memory}. A reliable computational framework arguably has more significance for non-Markov processes as memory dependence in general precludes the application of analytical methods. An implementation of cloning for memory-dependent processes can be found in~\cite{cavallaro}.

This paper is concerned with leveraging the efficiency of machine learning in large-deviation computations. From a broader perspective, there has been an increasing demand in statistical physics, condensed matter and active matter communities for tools that inherit the powerful mechanisms for pattern matching and data generation provided by machine learning \cite{thermodynamics_ML,Phases_of_Matter_ML,quantum_state_tomography,Active_matter_ML}. However, out of the many paradigms of machine learning, we are particularly interested here in reinforcement learning, an unsupervised framework for solving optimal control problems \cite{sutton}. This unsupervised character is the primary advantage of reinforcement learning and has already spawned learning-based control methods for statistical physics applications where training data is not readily available   \cite{Active_matter_ML,condensed_matter_RL,RL_for_quantum_thermodynamics}.  Moreover, as we shall see later in this work, it is well-known that SCGF computation can be formulated as an optimal control problem \cite{chetrite}, making reinforcement learning an attractive approach for large-deviation analysis. Indeed, frameworks based on reinforcement learning have already been developed for Markov systems with successful demonstrations on various models \cite{rose, diffusive,gillman2022reinforcement, Whitelam_2020,ml_scgf_touchette}.

 In this work, we extend the reinforcement learning  framework in \cite{rose} to obtain the SCGF in non-Markov systems, thus offering a potential new tool for understanding the effects of memory on rare events.  The paper is structured as follows. Section 2 introduces the nuances of memory dependence in stochastic systems via a discussion on semi-Markov processes where memory is modeled using nonexponential waiting-time distributions. Section 3 gives a brief overview of large deviations in nonequilibrium systems with a special focus on the optimal control approach used in many computational methods. Section 4 forms the core of this paper where we discuss the different components of the reinforcement learning  architecture. Section 5 demonstrates large-deviation applications on potentially interesting semi-Markov systems. Section 6 provides a summary and some perspectives for future work.  Appendix \ref{hmm} outlines a hidden-Markov analytical approach for checking some of the computational results. Appendix \ref{robustness} discusses the robustness of neural reinforcement learning with respect to the choice of key hyperparameters in the machine learning model, while appendices \ref{RNN method} and \ref{implementation} give various technical details of our implementation.

\section{Memory-dependent stochastic processes }

We introduce here the theoretical background for the simulation and analysis of memory-dependent trajectories, central to the computational approach discussed in this work. Our focus throughout will be on semi-Markov systems, described in this section in some detail. The semi-Markov assumption allows for the study of memory effects in a tractable setting and is often used to model real-world systems \cite{bio_semi_Markov,book_semi_Markov_applications,book_semi_Markov_applications_finance}. The study of semi-Markov systems is thus a physically-relevant step towards further generalization.    

\subsection{Semi-Markov trajectories}

An accessible starting point for studying memory dependence in stochastic-systems is the semi-Markov process, which generalizes the familiar Markov process by introducing time-dependent rates corresponding to nonexponential waiting times that are renewed each time an event occurs. A simple yet illustrative example of such a processes is the semi-Markov continuous-time random-walk (CTRW) \cite{ctrw_semi_Markov,semi_ctrw_asymptotics}. The model extends the memoryless CTRW \cite{CTRW} by  assuming that the time spent at a lattice point is governed by a nonexponential distribution and the time spent may also determine the transition probabilities. We use this CTRW example in discussions throughout this section and to showcase our computational method in section 4.6. 

An elegant mathematical formalism for the analysis of the general and asymptotic solutions of semi-Markov  processes (and some other non-Markov processes) is provided by the corresponding generalized Master equation, see \cite{semi_markov,semi_ctrw_asymptotics,GME_hauss}. However, here we restrict our attention to the trajectory formulation for semi-Markov processes, as described for example in \cite{andrieux-gaspard}. 

We examine here the continuous-time path evolutions under semi-Markov dynamics of configurations $x$  in a discrete state space $\mathcal{S}$. Such paths can be constructed using probability densities $\psi_{x',x}(\tau)$, each describing a transition from a given configuration $x$ to a new configuration $x'$ in the infinitesimal waiting-time interval $[\tau, \tau+ \mathrm{d}\tau)$. The densities follow the normalizing condition $\sum_{x'}  \int_{0}^{\infty} \psi_{x',x}(\tau)=1$, where the subscript indicates the sum is over all allowed transitions from state $x$. The time spent in a state $x$ is given by the \textit{waiting-time distribution}

\begin{equation}
\label{wt_formal}
\psi_{x}(\tau) =  \sum_{x'} \psi_{x',x} (\tau),
\end{equation}
with the associated \textit{survival function}

\begin{equation}
\varphi_{x}(\tau) = \int_{\tau}^{\infty} \psi_{x}(t) \mathrm{d}t.
\end{equation}

The waiting-time distributions can be used to express the path transitions in a convenient form for simulations:

\begin{equation}
\label{sim}
\psi_{x',x}(\tau) =  p(x'|x,\tau) \psi_{x}(\tau).
\end{equation}
Here  $p(x'|x,\tau)$ is a mass-function describing the probability of choosing a target configuration $x'$ for the jump out of $x$, having waited at that configuration for time $\tau$. As shown, the jump probabilities in semi-Markov processes in general depend on the waiting time $\tau$, violating \textit{direction-time independence}, a necessary (but not sufficient) condition for equilibrium in semi-Markov processes \cite{chari_94}. Furthermore, we note that \eqref{sim} can instead be written in terms of time-dependent process rates $\beta_{x',x}(\tau)$ as 

\begin{equation}
\label{rates}
\psi_{x',x}(\tau) = \beta_{x',x}(\tau) \varphi_{x}(\tau).
\end{equation}
Choosing exponential waiting-time distributions gives constant rates and the reduction to the Markov case.

The probability density of realizing a trajectory $\omega_{t_{0}}(t)$, starting at time $t_{0}$ in state $x_{0}$ and terminating at time $t$ in some state $x_n$  after $n$ observed transitions, in the infinitesimal  intervals $[t_k,t_{k}+\mathrm{d}t)_{k \in \{1,2,\cdots,n\}}$, may be written as

\begin{multline}
\label{trajectory1}
 \sigma(\omega_{t_{0}}(t))= \varphi_{x_{n}}(t-t_{n}) \psi_{x_{n},x_{n-1}}(t_{n}-t_{n-1}) \times \cdots  \\ \times \psi_{x_2,x_1}(t_{2}-t_{1}) \Psi_{x_{1},x_{0}}(t_1-t_0) p_{0}(x_0,t_0),
\end{multline}
with the condition, $t_{0}< t_{1}<t_{2}< \cdots <t_{n}< t$. Here $p_{0}(x_0,t_0)$ and $\Psi_{x_{1},x_{0}} (t_1-t_{0})$ describe the probability of finding the process in configuration $x_0$ at time $t_0$ and the density of the trajectory up to the first state transition. The term, $\varphi_{x_{n}}(t-t_{n})$ represents the survival function indicating that no jumps out of $x_{n}$ occur until the final observation time $t$. 

The description of initial conditions and the corresponding distributions require careful attention in systems with memory. In fact, for non-Markov, processes it is not sufficient to describe initial conditions just with the starting state $x_0$ and time $t_0$ but one also needs information on the history of the process before the observation begins. For instance, for the semi-Markov trajectory description in \eqref{trajectory1},  $p_{0}(x_0,t_0)$ and $\Psi_{x_1,x_0}(t_1-t_0)$ also depend on how long the process has been in the state $x_0$ before $t_0$. Thus, to describe these distributions, we require the history of the trajectory before $(x_0,t_0)$, or alternatively, some average over all such histories.  To elaborate on a particular case, if we know that the process has been in the state $x_0$ for some time $\tau_{h}$ before the initial time $t_0$, the initial density takes the form, $\Psi_{x_{1},x_{0}}(t_{1}-t_{0})= \psi_{x_1,x_0}(t_{1}-t_{0}+\tau_h)/ \varphi_{x_{0}} (\tau_{h})$. Our notation suppresses these details in favor of retaining the general form of the distributions, without reference to history before $t_{0}$.

We may then use the trajectory probability densities \eqref{trajectory1} for further analysis of the dynamics. For example,  the probability of ending up in a state $x \in \mathcal{S}$ at time $t$ starting from $x_{0}$ at time $t_0$  can be calculated as
\begin{multline}
p(x,t,x_0,t_0)= \delta_{x,x_0} \Phi_{x_0}(t-t_{0}) p(x_0,t_0) \\ +\sum_{n=1}^{\infty}  \int_{t_{0}}^{t} \int_{t_{1}}^{t} \dots \int_{t_{n-1}}^{t} \mathrm{d}t_{1} \mathrm{d}t_{2} \dots \mathrm{d}t_n \sum_{x_1...x_{n}} \delta_{x,x_n}  \sigma(\omega_{t_{0}}(t)). 
\end{multline}
Here $\Phi_{x_0}(t-t_{0})$ is the survival function for the case where the process remains in state $x_{0}$ until the time $t$. Again, for a semi-Markov process which has been in the state $x_{0}$ for a time $\tau_{h}$ before $t_{0}$, we have $\Phi_{x_{0}}(t-t_0) = \varphi_{x_{0}}(t-t_{0}+\tau_h)/ \varphi_{x_0}(\tau_h)$.

Using the form given in \eqref{sim}, there is also an equivalent representation of the trajectory probability  density \eqref{trajectory1} in terms of waiting-time distributions and probability mass functions,
\begin{multline}
\label{trajectory2}
 \sigma(\omega_{t_0}(t))= \varphi_{x_{n}}(t-t_{n}) p(x_{n}|x_{n-1},t_{n}-t_{n-1}) \psi_{x_{n-1}}(t_{n}-t_{n-1}) \times \cdots \\  \times p(x_2|x_1,t_{2}-t_{1})\psi_{x_1}(t_{2}-t_{1}) \Psi_{x_{1},x_{0}}(t_1-t_0) p_{0}(x_0,t_0).
\end{multline}
This allows simulation via a modification of the Monte Carlo method due to Gillespie \cite{GILLESPIE1976403}. At a generic step $k$ of the algorithm,  a  waiting-time $\tau_{k}$ from $\psi_{x_{k}}(\tau)$ and a state transition $x_{k} \to x_{k+1}$ according to $p(x_{k+1}|x_{k},\tau_{k})$ are sampled. The procedure is repeated until a terminal time $t$ is reached.

As a concrete example, we return to the CTRW model mentioned at the beginning of this subsection. In particular, we consider a bidirectional semi-Markov CTRW on a discrete lattice. To avoid complications with infinite state space, we assume a finite lattice of size $L$ with periodic boundaries such that the state updates are modulo-$L$. 

The dynamics is realized by two competing nonexponential probability densities $\psi^{+}(\tau)$ and $\psi^{-}(\tau)$ triggering the forward and backward jumps respectively. The corresponding survival functions are denoted $\varphi^{+}(\tau)$ and $\varphi^{-}(\tau)$.    
We assume here that both distributions are renewed after a jump out of a site occurs, i.e., all memory is lost after a transition event. One may  imagine the above densities are associated with two competing clocks at each lattice site.\footnote{It is worth noting that, in this model, attaching the clocks to the site is equivalent to attaching the clocks to the particle. However, this may not be true for more complicated models.} The two clocks count down to the forward and backward transitions respectively and the transition associated with the winning clock (the one which rings first) is triggered. We then have the forward and backward transition densities
\begin{equation}
\psi_{x+1,x}(\tau) = \psi^{+}(\tau) \varphi^{-}(\tau), \quad  \psi_{x-1,x}(\tau) =\psi^{-}(\tau)\varphi^{+}(\tau).
\end{equation}
The first density gives the instance when the forward clock wins and the backward clock loses, while the second density gives the opposite scenario. From \eqref{wt_formal}, the overall waiting-time distribution at a specific state of the process is then given by the combination

\begin{equation}
\label{ctrw_wtd}
\psi_{x}(\tau)=\psi^{+}(\tau) \varphi^{-}(\tau) + \psi^{-}(\tau) \varphi^{+}(\tau).
\end{equation}
The overall system survival function  is $\varphi_{x}(\tau)=\varphi^{+}(\tau) \varphi^{-}(\tau)$, corresponding to no transitions occurring in a time interval of length $\tau$ at state $x$. Note, the process is homogeneous and the waiting-time distribution is independent of the state $x$. This means that the transitions at each lattice site of the CTRW ring all have the same clock structure described above. Finally, using \eqref{rates}, we may write the transition rates out of a state

\begin{equation}
\beta^{+}(\tau)=\frac{\psi^{+}(\tau)}{\varphi^{+}(\tau)} , \quad
\beta^{-}(\tau)=\frac{\psi^{-}(\tau)}{\varphi^{-}(\tau)},
\end{equation}
and the associated probability mass functions
\begin{equation}
p(x+1|x,\tau) = \frac{\beta^{+}(\tau)} {\beta^{-}(\tau)+{\beta^{+}(\tau)}}, \quad  p(x-1|x,\tau) = \frac{\beta^{-}(\tau)}{\beta^{+}(\tau)+\beta^{-}(\tau)}.
\end{equation}
The above quantities can then be utilized for trajectory simulations via \eqref{trajectory2}.
We will come back to this semi-Markov CTRW example to demonstrate results from our computational method in section 4.6.

\subsection{Long-time limit}

The trajectory perspective is useful for studying the statistical properties of various physical quantities that provide specific insights about the system. We are especially interested in the behavior of quantities over extremely long timescales, analogous to the \textit{thermodynamic limit} in equilibrium statistical physics. In this work we focus on stochastic dynamics that lead to time-homogeneous processes that lead to a well-defined stationary state. In the time-homogeneous Markov case, this condition is met when the ergodic property holds \cite{ctmc_ergodicity}, but non-Markov processes may need to satisfy further criteria. For semi-Markov systems, with rates in the extended state space of configurations and waiting times that do not depend  on the process time, then a stationary state is guaranteed if, along with the ergodic property, the waiting-time distributions have finite moments \cite{ctrw_semi_Markov, semi_ctrw_asymptotics}.

An important quantity of interest in nonequilibrium statistical physics is the time-integrated current $J(\omega(t))$, which counts directed jumps in the trajectory $\omega(t)$; here and in the following we often suppress the dependence on initial time $t_0$ in the notation for the trajectory. For example, in the CTRW case, one can define a current by counting all particle
jumps across the system lattice with the current incremented or decremented based on whether a forward or a backward jump occurs.  Alternatively, one can choose to count the current across particular bonds or a subset of bonds as we shall see in the multi-particle examples of section 5.2. These definitions are  closely related and we will use the symbol $J(\omega(t))$ for a generic current with the time antisymmetric property. Note that if the system is at \textit{equilibrium}, the  time averages of all currents in the system are zero. We define equilibrium here as process symmetry under time reversal, i.e., time-reversed trajectories being statistically indistinguishable from forward trajectories. In a Markov system, the statement that the currents between \textit{all} pairs of states are zero is equivalent to the detailed-balance condition which is well-known to be necessary and sufficient for equilibrium.\footnote{This implicitly requires that, if a transition $x\to x'$ is possible, then the reverse transition $x' \to x$ is also possible.} However, in the semi-Markov case,  zero currents are necessary but not sufficient for equilibrium. To be precise, for a semi-Markov system to be in equilibrium it must obey both the principle of direction-time independence, such that the jump probabilities are independent of waiting times, and a detailed-balance condition on those jump probabilities \cite{chari_94, conditions_for_reversibility}. As an example of a nonequilibrium semi-Markov system with zero currents, one may easily imagine a continuous-time random walk where the waiting times for forward and backward jumps have different distributions but the same mean.

For out-of-equilibrium systems, the long-time behavior of currents is often more interesting. When one has a nonequilibrium stationary state   \cite{NESS, NESS_2}, time-intensive currents $j=J(\omega(t))/{t}$ generically converge to typical (nonzero) values with well-defined moments for large but finite times. Fluctuations away from the typical values correspond to exponentially rare events. The study of such rare events is the main subject of this paper.
 
 As rare events by definition occur far away from typical trajectories, observing them in simulations via the Gillespie method becomes very inefficient. One idea towards developing an efficient simulation procedure is by \textit{tilting} or exponential reweighting of trajectory ensembles as suggested by the thermodynamic partition function,
 
\begin{equation}
\label{partition_function}
 Z(s,t)  = \int e^{sJ(\omega(t))} \sigma(\omega(t))\mathrm{d \omega(t)}.
\end{equation}
This reweighting procedure falls under the purview of importance sampling methods 
framework using the tilting procedure 
for simulating rare events is discussed in section 3.

Non-Markov theory, of course, extends beyond the simple semi-Markov processes considered above. For instance, process dynamics could depend on one or more hidden variables other than the waiting times. We believe that the method shown in this paper is applicable to all processes that relax to well-defined stationarity in the state space including the hidden variables; this might include processes with weakly time-inhomogeneous rates (which become constant in the long-time limit). However, for non-Markov processes with time-inhomogeneous dynamics in the long-time limit, significant changes to the computational method would be required. A notable class of such processes are systems in the spirit of the elephant and Alzheimer's random walks \cite{ERW_shutz, Kenkre_ERW}, where the dynamics depends on the time-averaged current; some related applications can be found in \cite{amnesia_alzhimer_application, Gilchrist_ribosome,sharma_ribosome}. Large fluctuation analysis for models of such type may be found, for example, in \cite{Harris_erw, Harris_2015, jack_harris}. Intuitively, dynamics on an extended state space including time-averaged current $j$ is not time homogeneous since a given increment in position space has a decreasing effect on $j$ as time increases. For models of this kind ergodicity is often broken; the extension of our method to nonergodic processes is left to future work.

\section{Current fluctuations} 

\subsection{Large deviations}
In Markov processes, the statistics of time-averaged quantities such as  currents usually follow a large-deviation principle loosely stated as

\begin{equation}
\label{LDP}
\mathrm{Prob} \left( \frac{J(\omega(t))} {t} =j \right) \sim e^{-tI(j)},
\end{equation}
where the symbol $\sim$ indicates logarithmic equality in the long-time limit. Here, $t$ is sometimes called the \textit{speed} of the large deviation. In non-Markov processes, the large-deviation principle may be modified \cite{jack_harris, Harris_2015} or even not exist at all. However, for a class of non-Markov systems, a large-deviation principle with linear speed does still hold and we restrict ourselves in this paper to such cases.    

The rate function  $I(j)$ in the large-deviation principle  governs the occurrence probability of rare events, i.e., rare currents in the present case, and is a central quantity in  nonequilibrium analysis. However, the rate function is in general difficult to obtain and is often computed indirectly via Legendre-Fenchel transformation of the SCGF, as prescribed by the G\"artner-Ellis theorem:

\begin{equation}
 \label{GE}
 I(j) = \max_{s} \{ sj - \lambda(s) \}.          
 \end{equation}
 
From a thermodynamic standpoint, the SCGF is related to the partition function in \eqref{partition_function} and is formally defined as 

\begin{equation}
 \label{SCGF1}
 \lambda(s) = \lim_{t\to \infty} \frac{1}{t}\ln Z(s,t) = \lim_{t \to \infty} \frac{1}{t} \ln   \left \langle     e^{s t j}  \right \rangle  _{\sigma(\omega(t))},  
 \end{equation}
where the average is over trajectory probabilities $\sigma(\omega(t))$ of the process. 

\subsection{An optimal control approach}
 
In the Markov case, the SCGF is the principal eigenvalue of the tilted (exponentially reweighted) form of the rate matrix governing the dynamics and for simple examples can be calculated analytically.  Beyond these simple Markov cases and certainly in the non-Markov context, the SCGF may take a more complicated form, making analytical progress prohibitively difficult.  Moreover, complications such as phase transitions may arise even in Markov systems \cite{phase_transitions,Rosemary_zrp}. Large deviations in many cases can only be probed numerically, highlighting the importance of efficient computational techniques. The broad objective of most techniques referenced in this work is to obtain the SCGF through trajectories of an \textit{alternate} or \textit{modified} dynamics whose typical behavior corresponds to the large fluctuations of the original dynamics. The probability of a trajectory under the alternate dynamics is then given as  
\begin{equation}
\sigma^{w}(\omega(t)) = e^{sJ(\omega(t))} \frac{\sigma(\omega(t))}{Z(s,t)},
 \end{equation}
where the exponentially weighted trajectory probabilities are  normalized by the partition function $Z(s,t)$.  For Markov systems, the alternate dynamics is obtained via the Doob transform, see e.g.\ \cite{chetrite}, which is utilized in cloning algorithms \cite{giardina,Lecomte_2007}. 

A more general way of obtaining the alternate dynamics is by  using a variational approach, as in \cite{Jacobson_2019}, applied over a parameterized model representing  the alternate dynamics. In particular, as shown in \cite{chetrite} the alternate dynamics is estimated via the minimization of the Kullback-Liebler  divergence (KLD)
\begin{align}
   \label{D_KL}
    D_{\mathrm{KL}}(\sigma^{\theta}(\omega(t)),\sigma^{w}(\omega(t))) & =   \left \langle     \ln \frac{\sigma^{\theta}(\omega(t))} {\sigma^{w}((\omega(t))}   \right \rangle  _{\sigma^{\theta}(\omega(t))}  \nonumber \\ 
    & =   \left \langle     -sJ(\omega(t)) + \ln\frac{\sigma^{\theta}(\omega(t))}{\sigma(\omega(t))}    \right \rangle  _{\sigma^{\theta}(\omega(t))} + \ln Z(s,t).
\end{align}
Here, $\sigma^{\theta}(\omega(t))$ is the control distribution parametrized by $\theta$.  Time averaging, taking the long-time limit and using the fact that the KLD is by definition non-negative, leads to the following bound involving the SCGF:  
\begin{equation}
\label{bound}
\lim_{t \to \infty} \frac{1}{t}   \left \langle     sJ(\omega(t)) - \ln\frac{\sigma^{\theta}(\omega(t))}{\sigma(\omega(t))}   \right \rangle  _{\sigma^{\theta}(\omega(t))} \leq \lambda(s).
\end{equation}
In other words, the SCGF is bounded from below by the quantity on the left-hand side. This bound is critical to variational methods and in section 4.2 we demonstrate that it can be used with reinforcement learning for the efficient computation of the SCGF in semi-Markov systems.

\section{Computational framework}

In this section we describe how to extend the reinforcement learning scheme by Rose, Garrahan and collaborators  \cite{rose,diffusive,gillman2022reinforcement}.   
The reinforcement learning framework in their works aims to conduct a systematic and iterative search for atypical trajectories through the adaption of appropriately parameterized dynamics. In the case even of semi-Markov systems this becomes very complicated and may involve a large number of parameters. However, we show in this section that using multilayer  neural networks, makes such complicated optimizations possible.

\subsection{The reinforcement problem}

At its core, reinforcement learning is an optimal control technique that uses environmental feedback in the form of rewards to evaluate the quality of actions leading to a trajectory step \cite{sutton,Bartsekas_DP_vol1}. In recent years, combining reinforcement learning with neural architectures has given rise to a number of diverse and powerful techniques \cite{dqn, a3c, PPO}.       

The reinforcement learning system used in the present context takes inspiration from the soft actor-critic framework \cite{sac}. The \textit{actor-critic} structure, as the name suggests, has an \textit{actor} that searches the space of parameterized dynamics known as policies while a separate \textit{critic} evaluates the policies by analyzing their effect on the environment. Each of these systems are then improved separately by using gradient-based techniques. Another vital characteristic of the framework used here is the entropy-regulation of the rewards \cite{Todorov,sac}. A KLD term akin to \eqref{D_KL} regulates the rewards and conditions the learning towards the desired dynamics.

We consider the reinforcement problem on the extended state space that includes process memory. In particular, the reinforcement learning problem in the semi-Markov domain is posed as a decision process in the joint space of state configurations and waiting times. The decision process in the extended state space is Markovian and can be solved recursively.  Related ideas appear in \cite{SUTTON_smdp, Bartsekas_DP_vol1,Doshi_1979,Schmidhuber,WHITEHEAD_NM}.

To construct a semi-Markov decision process, we examine here the transitions $ (x',\tau') \to  (x'', \tau'')$ in the extended state space of configurations and waiting times, $ \mathcal{S} \times \mathcal{T}$. In words, the process having waited at a configuration $x'$ for a time $\tau'$ chooses the new state $x''$ and the time $\tau''$ it must wait there until a new decision is made. From a trajectory standpoint, we assume that the process departs  states $x'$ and $x''$ at trajectory times $t'$ and $t''$. We now return to  denoting a trajectory  starting at time $t_0$ and ending at time $t$ as $\omega_{t_0}(t)$. Then the piece of the trajectory $\omega_{t'}(t'')$ is generated  by the density
\begin{equation}
\sigma^{\theta^{p},\theta^{q}}(\omega_{t'}(t'')) = \pi_{x''}^{\theta^{q}}(\tau'')\pi^{\theta^{p}}(x''|x',\tau').
\end{equation}
Here $\theta^p$ and $\theta^q$ represent two independent sets of parameters labeled with superscripts $p$ and $q$. These separately construct adaptable sequential stochastic policies  $\pi^{\theta^{p}}$ and $\pi^{\theta^{q}}$ controlling the jump to a new state and the new waiting-time respectively. As the parameter description allows policies to be independently updated, the learning simplifies \cite{orthogonal}. The construction of this two-policy structure with  neural networks is one of the key innovations of our method and will be explained in detail in section 4.3. 

In general, the generation system described above can be used to simulate semi-Markov trajectories $\omega_{t_0}(t)$ with densities
\begin{multline}
\label{decision_process}
\sigma^{\theta^{p},\theta^{q}}(\omega_{t_0}(t)) 
 =  \varphi_{x_{n}}(t-t_n)  \pi^{\theta^p}(x_n|x_{n-1},\tau_{n-1}) \times \cdots \\  \times\pi^{\theta^q}_{x_2}(\tau_2) \pi^{\theta^p}(x_2|x_1,\tau_1) \pi^{\theta^q}_{x_1}(\tau)\Psi_{x_1,x_0}(t_{1}-t_{0}) p_0(x_0,t_0),
\end{multline}
analogous to the form given in \eqref{trajectory2}. Similar to \eqref{trajectory2},  $\Psi_{x_{1},x_{0}}(t_1-t_0)$ and $\varphi_{x_{n}}(t-t_{n})$ are special and depend on the initial and terminal trajectory times selected. We do not explicitly define separate policies for these densities as their effect on the trajectory density is small in the long-time limit, when the waiting-time distributions are assumed to have finite moments. We use this argument later while defining the rewards in the next section. Although we focus here on semi-Markov systems, the technique presented can, in principle, be applied to other hidden-variable systems with minimal modifications. 

To apply reinforcement learning to the extended state space described above, we next expand the general actor-critic framework to a multi-agent setting \cite{multi_agent_RL} for simultaneously learning the two policies.

\subsection{Policy gradient}

We extend the procedure used in \cite{rose} to construct an actor-critic reinforcement algorithm for semi-Markov systems. We begin by rewriting the KLD of \eqref{D_KL} in terms of the two-policy decision process  given in \eqref{decision_process}:

\begin{align}
\label{D_kl_2}
  D_{\mathrm{KL}}(\sigma^{\theta^{p},\theta^{q}},\sigma_{w}) & =    \left \langle     \ln \frac{\sigma^{\theta^{p},\theta^{q}}(\omega_{t_0}(t))}{\sigma^{w}(\omega_{t_0}(t))}  \right \rangle  _{\sigma^{\theta^{p},\theta^{q}}} \nonumber \\ & =  \left \langle     -sJ(\omega_{t_0}(t))+   \ln{\frac{\sigma^{\theta^{p},\theta^{q}}(\omega_{t_0}(t))}{\sigma(\omega_{t_0}(t))}}  \right \rangle  _{\sigma^{\theta^{p},\theta^{q}}} + \ln{Z(s,t)}. 
\end{align}
This further allows rewriting of the bound in \eqref{bound} as 
\begin{equation}
\lambda(s) \geq  \lim_{t \to \infty} \frac{1}{t}   \left \langle     sJ(\omega_{t_0}(t)) - \frac{\sigma^{\theta^{p},\theta^{q}}(\omega_{t_0}(t))}{\sigma(\omega_{t_0}(t))}   \right \rangle  _{\sigma^{\theta^{p},\theta^{q}}}.
\end{equation}
We may then define the return over a  specific finite trajectory,    
\begin{equation}
\label{return}
R(\omega_{t_0}(t))=   sJ(\omega_{t_{0}}(t)) -\ln{\frac{\sigma^{\theta^{p},\theta^{q}}(\omega_{t_0}(t))}{\sigma(\omega_{t_0}(t))}}, 
\end{equation}
to arrive at  
\begin{equation}
\lambda(s) \geq \max_{\theta^{p},\theta^{q}} \{\lim_{t \to \infty} \frac{1}{t}   \left \langle     R(\omega_{t_{0}}(t))  \right \rangle  _{\sigma^{\theta^{p},\theta^{q}}}\}
\end{equation}
as an optimization problem for the SCGF.
 Assuming $n$ transitions, a recursive optimization algorithm can be constructed  by observing that the trajectory returns can broken down into piecewise rewards:
\begin{equation}
R(\omega_{t_0}(t))  =\sum_{k=1}^{n}r_{k} (x_{k},\tau_{k},x_{k-1},\tau_{k-1}). 
\end{equation}
For a generic transition, $(x_{k-1},\tau_{k-1}) \to (x_{k}, \tau_{k})$, the reward may be defined as
\begin{equation}
\label{reward}
r_{k}(x_{k},\tau_{k},x_{k-1},\tau_{k-1}) = sJ(x_k,x_{k-1})-\ln \frac{\pi_{x_k}^{\theta^{q}}(\tau_k) \pi^{\theta^{p}}(x_{k}|x_{k-1},\tau_{k-1})}{\psi_{x_k}(\tau_k)p(x_{k}|x_{k-1}, \tau_{k-1})},
\end{equation}
where $J(x_k,x_{k-1})$ represents the change in the integrated current in the transition from $x_{k-1}$ to $x_k$. In a bidirectional random walk, for example,  the current is incremented by one for a forward jump and decremented by one for a backward jump.

In continuous time, for fixed initial and final times $t_0$ and $t$, the number of transitions is itself a random variable. The initial and terminal rewards do not follow the form given in \eqref{reward}. In practice, we ignore these technical considerations, noting that the contribution of initial and final rewards to the return is small for long trajectories, at least when the waiting-time distributions have finite moments. 

We are interested here in the quantity $\left \langle  R(\omega_{t_0}(t))  \right \rangle _{\pi^{\theta^q},\pi^{\theta^p}}$, which is the average return over the trajectories simulated by fixing the policies $\pi^{\theta^q}$ and $\pi^{\theta^p}$ (taking into account the variable number of transitions).
Intuitively, one can at this point directly optimize the return by applying a perturbative method. Computing the gradient of \eqref{D_kl_2}  separately with respect to parameters $\theta^{p}$
and $\theta^{q}$ gives, 

\begin{align}
\label{policy_gradients}
& \nabla_{\theta^{p}} D_{\mathrm{KL}}(\sigma^{\theta^p,\theta^q}|\sigma^{w}) = - \left \langle      \sum_{k=1}^{n}     \left(R(\omega_{t_k}(t))-\bar{b}\right) \nabla_{\theta^p} \ln \pi^{\theta^{p}}(x_k|x_{k-1},\tau_{k-1})  \right \rangle  _{\pi^{\theta^p},\pi^{\theta^{q}}}  \nonumber \\ 
& \nabla_{\theta^{q}} D_{\mathrm{KL}}(\sigma^{\theta^p,\theta^q}|\sigma^{w})=-  \left \langle      \sum_{k=1}^{n}       \left( R(\omega_{t_k}(t)-\bar{b}) \right)  \nabla_{\theta^{q}}\ln \pi^{\theta^{q}}_{x_k}(\tau_k)   \right \rangle_{\pi^{\theta^{p}},\pi^{\theta^{q}}}.
\end{align}
Here the averages are taken over trajectories obtained by following the parameterized policies. The term $\bar{b}$ is an optional \textit{baseline}, with possible dependence on the state, introduced to control the variance of the gradient updates. The derivation of \eqref{policy_gradients} is similar to the derivation of the policy gradient in \cite{rose}. 
The above suggests a computation method using averaged returns over   Monte Carlo trajectories as  seen in policy-gradient methods \cite{policy_gradient}. 
However, computation of the above gradient is very inefficient for large trajectories. 

\subsection{Value gradient}

A reliable strategy to minimize the variance in the policy gradients  is to define a  critic evaluating the \textit{value function} pertaining to the system state space \cite{sutton,konda, a3c}.  A value function maps states to the expected future returns. From an optimal control perspective, this gives the \textit{value} of including a state in the controlled dynamics. We define a value function as the expected return over all trajectories departing from the joint state $(x',\tau')$ at some time $t'$ and terminating at later time $t$. In the case of rare-event analysis, the value of the state $(x',\tau')$ may be understood as its contribution to the rare trajectory. We write the value function

\begin{equation}
V_{\pi^{\theta^{p}},\pi^{\theta^{q}}}(x',\tau',t') =    \left \langle   R(\omega_{t'}(t))  \right \rangle  _{\pi^{\theta^{p}},\pi^{\theta^{q}},x',\tau'},
\end{equation}
where the average is over trajectories with departures from a particular state $x'$ occurring in the infinitesimal interval $[t',t'+dt)$ after a waiting-time $\tau'$ elapses. The value function follows the recursive Bellman principle \cite{BERTSEKAS_multi_agent_value_iteration,SUTTON_smdp}

\begin{equation}
\label{belman}
V_{\pi^{\theta^{p}},\pi^{\theta^{q}}}(x',\tau',t') =   \left \langle     V_{\pi^{\theta^{p}},\pi^{\theta^{q}}}(x,\tau,t'+\tau) + r(x,\tau, x',\tau')  \right \rangle  _{\pi^{\theta^{p}},\pi^{\theta^{q}},x', \tau'}.
\end{equation}
Of course, the value function is not known initially and must also be evaluated while policy improvements are performed. Towards this end, the function is parameterized with $\phi$ and may be approximated via the Bellman equation \eqref{belman} to give
\begin{align}
\delta^{\phi}_{td}(x_{k},\tau_{k}, x_{k-1},\tau_{k-1},t_{k}) = V_{\phi}( & x_{k},\tau_{k},t_{k}  +\tau_{k}) \nonumber \\ & +r(x_{k},\tau_{k},x_{k-1},\tau_{k-1}) -V_{\phi}(x_{k-1},\tau_{k-1},t_{k}). 
\end{align}
The term $\delta^{\phi}_{td}$ is known as the \textit{temporal-difference} error \cite{Sutton_temporal_diff} and can be minimized using the gradient-descent method via a mean-squared loss. A \textit{semi-gradient} method \cite{sutton} is used to simplify computations, fixing the parameters of the first value function to provide a \textit{target} for the second value function. Given that we have parameters $\phi = \tilde{\phi}$ at a particular step of the algorithm, the target approximates the right-hand side of the Bellman equation at $\tilde{\phi}$. We then have the loss function

\begin{align}
\label{value_loss}
L_{v}(\tilde{\phi},\phi)  = &\frac{1}{2}     \left \langle     \sum_{k=1}^{n} \left( V_{\tilde{\phi}} (x_{k},\tau_{k},t_{k}+\tau_{k})+r(x_k,\tau_k,x_{k-1},\tau_{k-1}) \right. \right. \nonumber \\  & \phantom{.........} \left. \left.  - V_{\phi}(x_{k-1},\tau_{k-1},t_{k})|_{\phi=\tilde{\phi}} \right)^{2}   \right \rangle_{\pi^{\theta^{p}},\pi^{\theta^{q}}},
\end{align}
and computing the gradient with respect to  $\phi$ gives

\begin{align}
&\nabla_{\phi}L_{v}(\tilde{\phi},\phi)  \nonumber \\ & =      \left \langle     \sum_{k=1}^{n}  \left( V_{\tilde{\phi}} (x_{k},\tau_{k},t_{k}+\tau_{k}) +r(x_k,\tau_k,x_{k-1},\tau_{k-1}) \right. \right.  \nonumber \\ & \left. \phantom{.....\sum_{k=1}^{n}} \quad  \left. -V_{\phi}(x_{k-1},\tau_{k-1},t_{k})|_{\phi=\tilde{\phi}} \right) \nabla_{\phi}V_{\phi}(x_{k-1},\tau_{k-1},t_{k}) \right\rangle_{\pi^{\theta^{p}},\pi^{\theta^{q}}} \nonumber \\  & = 
 \left \langle     \sum_{k=1}^{n}\delta^{\phi}_{td} (x_{k},\tau_k, x_{k-1},\tau_{k-1},t_{k}) |_{\phi = \tilde{\phi}} \nabla_{\phi}V_{\phi}(x_{k-1},\tau_{k-1},t_{k})    \right \rangle  _{\pi^{\theta^{p}},\pi^{\theta^{q}}}.
\end{align}
Since the target estimates the future returns we use this in \eqref{policy_gradients}. Applying the value approximation of the current-state as the baseline  we may write
\begin{align}
  \left \langle    R(\omega_{t_{k}}(t))   \right \rangle  _{\pi^{\theta^p},\pi^{\theta^{q}}} -\bar{b} & \approx  \left \langle     V_{\tilde{\phi}} (x_{k},\tau_{k},t_{k}+\tau_{k})+r(x_k,\tau_k,x_{k-1},\tau_{k-1})  \right \rangle  _{\pi^{\theta^p},\pi^{\theta^q}} \nonumber \\ &   \phantom{............} -V_{\phi}(x_{k-1},\tau_{k-1},t_{k}) |_{\phi=\tilde{\phi}}
\nonumber \\ & = \delta_{td}^{\phi} (x_{k},\tau_{k},x_{k-1},\tau_{k-1},t_{k}) |_{\phi= \tilde{\phi}},
\end{align}
giving, as in \cite{a3c, rose, diffusive}, the policy gradients in terms of the temporal-difference error:

\begin{align}
& \nabla_{\theta^{p}} D_{\mathrm{KL}}(\sigma^{\theta^p,\theta^q}|\sigma^{w}) \approx  -  \left \langle     \sum_{k=1}^{n} \delta_{td}^{\phi} (x_k,\tau_{k},x_{k-1},\tau_{k-1},t_{k}) \nabla_{\theta^p} \ln \pi^{\theta^{p}}(x_k|x_{k-1},\tau_{k-1})  \right \rangle_{\pi^{\theta^{p}},\pi^{\theta^q}}
\nonumber \\ 
& \nabla_{\theta^{q}} D_{\mathrm{KL}}(\sigma^{\theta^p,\theta^q}|\sigma^{w}) \approx -  \left \langle      \sum_{k=1}^{n} \delta_{td}^{\phi}(x_{k},\tau_{k},x_{k-1},\tau_{k-1},t_{k}) \nabla_{\theta^q}\ln \pi_{x_{k}}^{\theta^{q}}(\tau_k)   \right \rangle  _{\pi^{\theta^{p}},\pi^{\theta^{q}}}.
\end{align}

\subsection{Differential actor-critic}
\label{diff_act_crit}
In the long-time limit, applying the reinforcement learning method leads to very large returns which causes the value function to diverge. One way to resolve this issue is to consider a \textit{differential-reward} setting    \cite{sutton,Bartsekas_DP_vol1,Mahadevan1996}. 

Differential reinforcement learning makes use of ergodic properties that lead to time homogeneous stationary states. In the semi-Markov case, with finite waiting-time moments, time-homogeneity of the stationary dynamics is guaranteed in the extended state space of configurations and waiting-time distributions \cite{semi_ctrw_asymptotics}. This enables the application of differential reinforcement learning. A similar argument for the application of the method can be made when the state space is extended to include hidden variables in other non-Markov systems. 

We denote the average reward as
\begin{equation}
\bar{r}_{\theta^{p},\theta^{q}} =  \lim_{t \to \infty} \frac{1}{t}   \left \langle     R(\omega_{t_{0}}(t))  \right \rangle  _{\pi^{\theta^{p}},\pi^{\theta^{q}}},
\end{equation}
which gives the following \textit{differential-return} over a sample trajectory  
\begin{equation}
R_D(\omega_{t_0}(t)) 
=\sum_{k=1}^{n} \left(r(x_{k},\tau_{k},x_{k-1},\tau_{k-1})-\tau_{k}\bar{r}_{\theta_{\tau},\theta_{p}} \right),
\end{equation}
where we have assumed that the reward structure for the first step is the same as the others and neglected the subleading contribution from the piece of the trajectory after the last transition.
This allows a new definition of a convergent value function in terms of differential-rewards, 

 \begin{equation}
V(x_0,\tau_{0}) = \lim_{t \to \infty}   \left \langle     R_{D}(\omega_{t_{0}}(t))  \right \rangle  _{\pi^{\theta^{p}},\pi^{\theta^{q}}},
\end{equation}
which obeys the modified Bellman condition,
\begin{equation}
\label{modified_bellman}
V(x',\tau') =   \left \langle     V(x,\tau) + r(x,\tau,x',\tau') - \tau \bar{r}_{\theta^{p},\theta^{q}}    \right \rangle  _{\pi^{\theta^{p}},\pi^{\theta^{q}},x',\tau'}.
\end{equation}
The above gives a definition for the time average of the KLD [cf.\ $\eqref{D_kl_2}$], which holds for any transition $(x',\tau') \to (x,\tau)$: 

\begin{equation}
d_{\mathrm{KL}} = - \frac{1}{t}   \left \langle     V(x,\tau)+ R(\omega_{t_0}(t))- V(x',\tau')  \right \rangle  _{\pi^{\theta^p},\pi^{\theta^q}} + \frac{1}{t}\ln Z(s,t).
\end{equation}
Taking derivatives with respect to the parameters and averaging over the stationary states of the dynamics under $\pi^{\theta^p},\pi^{\theta^q}$  gives the computational gradient estimates
\begin{align}
&\nabla_{\theta^{p}} d_{\mathrm{KL}} =-  \left \langle     \delta_{dtd}(x,\tau,x',\tau') \nabla_{\theta^{p}} \ln \pi^{\theta^{p}}(x|x',\tau')  \right \rangle_{\pi^{\theta^{p}},\pi^{\theta^{q}}} \nonumber, \\
&\nabla_{\theta^{q}} d_{\mathrm{KL}}=-  \left \langle      \delta_{dtd}(x,\tau,x',\tau') \nabla_{\theta^q}\ln \pi^{\theta^{q}}(\tau |x)   \right \rangle  _{\pi^{\theta^{p}},\pi^{\theta^{q}}}.
\end{align}
Here  $\delta_{dtd}(x,\tau,x',\tau')$ is the \textit{differential temporal difference} error obtained from the modified Bellman equation \eqref{modified_bellman} and is given as

\begin{equation}
 \delta_{dtd}(x,\tau,x',\tau') =  V_{\phi}(x,\tau) + r(x,\tau,x',\tau') - \tau \bar{r}_{\theta^{p},\theta^{q}}  - V_{\phi}(x',\tau').
\end{equation}
As before, this may be optimized by using the value gradient obtained with an objective similar to \eqref{value_loss}. 

The differential-reward method gives a viable actor-critic reinforcement learning scheme for long times. The policy and value parameters $\{\theta^{p}, \theta^{q}, \phi \}$  are updated using the gradients described above. The updates can be made \textit{online} (during the algorithm run) while simulating a very long trajectory. At each step of the actor-critic algorithm, the reward is computed from the actions generated by the policies. The temporal difference is computed via the value function estimate, which is then used to update the  actors and the critic. The update size is controlled by choosing learning rates $\alpha_{\theta^{p}},\alpha_{\theta^{q}},\alpha_{\phi}$ (typically small values of the order $10^{-4}$). Finally, at the end of every step we also update the estimate for the time-averaged rewards with some learning rate $\alpha_{\bar{r}}$, i.e., at the $i^{th}$ step of the run we have

\begin{align}
\bar{r}^{i+1}_{\theta^{p},\theta^{q}} =\bar{r}^{i}_{\theta^{p},\theta^{q}} + \alpha_{\bar{r}} \delta_{dtd}.
\end{align}
The complete actor-critic reinforcement method is shown in Algorithm \ref{rl_algo}.
\begin{algorithm}
\caption{Two-policy differential actor-critic }\label{rl_algo}
\begin{algorithmic}

\State \textbf{Parameters} Initial and terminal times $t_0$ and $t$;  average reward  $\bar{r}_{\theta^{p},{\theta^{q}}}$; temporal difference $\delta_{dtd}$, learning rates $\alpha_{\theta^{p}}$, $\alpha_{\theta^{q}}$ $\alpha_{\phi}$ and $\alpha_{\bar{r}}$; parameters of the actor networks to learn policies $\pi^{\theta^{p}}$ and $\pi^{\theta^{q}}$; parameters of the critic network to learn value function $V_{\phi}$.  

\State \textbf{Initialize} Set $(x',\tau')$ to  $(x_{0},\tau_{0}+\tau_{h})$. $x_0$ is a random state in the configuration space. $\tau_h$ is the time spent in $x_0$ before $t_{0}$. $\tau_{0}$ is the time spent in $x_0$ after $t_{0}$ until the first departure.  $\tau_{0}+\tau_{h}$ is sampled from a uniform distribution. Initialize neural network parameters for the actors and the critic. Initialize starting time $t_0$ and starting step of the run $i=1$. 
\State $t_{1} \gets t_{0}+\tau_{0}$

\While {$t_{i}<t$}
\State Generate a transition $ x$ with $\pi^{\theta^{p}}(x|x',\tau')$
\State Generate a waiting-time for state $\tau$ with $\pi^{\theta^{q}}_{x}(\tau)$
 \State Compute $r(x,\tau,x',\tau')$
 \State Get values $V_{\phi}(x',\tau')$ and $V_{\phi}(x,\tau)$ from the critic
 \State $\delta_{dtd} \gets  V_{\phi}(x',\tau')+r(x,\tau,x',\tau')- \tau \bar{r}_{\theta^{p},\theta^{q}}-V_{\phi}(x',\tau')$

\State $\theta^{p} \gets \theta^{p} + \alpha_{\theta^{p}} \delta_{dtd} \nabla_{\theta^{p}}\ln{\pi^{\theta^{p}}(x|x',\tau'})$

\State  $\theta^{q} \gets \theta^{q} + \alpha_{\theta^{q}} \delta_{dtd} \nabla_{\theta^{q}}\ln{\pi^{\theta^{q}}_{x}}(\tau)$

\State $\phi \gets \phi + \alpha_{\phi} \delta_{dtd} \nabla_{\phi}V_{\phi}(x',\tau')$

\State $\bar{r}_{\theta^{p},\theta^{q}} \gets  \bar{r}{_\theta} +  \alpha_{\bar{r}} {\delta_{dtd}}$

\State   $x,\tau \gets x',\tau'$ 
\State $t_{i+1} \gets t_{i}+\tau$
\State $i \gets i+1$

\EndWhile
\end{algorithmic}
\end{algorithm}

We again direct readers unfamiliar with reinforcement learning to works \cite{sutton,sac,konda, a3c,multi_agent_RL} for an introduction and deeper explanation of the general actor-critic framework and the multi-agent setting. For an indication of how to visualize the two-policy gradients see \cite{orthogonal}, which explains gradient-based learning for the different but related problem of performing consecutive tasks.

\subsection{Neural architectures}
\label{Neural architectures}
Our multi-agent reinforcement learning architecture consists of two actors and a common critic. The components are constructed using  neural networks,
which have shown promise as representational frameworks and have now become the standard in modern reinforcement learning methods \cite{alphago,a3c,policy_gradient,PPO}. To help readers with a background in statistical physics, we provide here a short introduction to neural networks.

A neural network is a universal approximator that learns complicated functions \cite{HORNIK_NN,NN}. Structurally, the basic feed-forward neural network unit performs a weighted sum of the input and applies an \textit{activation} function such as the hyperbolic tangent to generate non-linearity. Then, several  layers of such units are stacked to create a connected graph, forming the multilayered feed-forward neural network. Mathematically, the  structure is a composition of functions that apply a complicated series of transformations to the input vector to generate an output. Thus, an $l$-\textit{layered} neural network with input $\mathbf{x}$ performs the weighted transformations, $\mathbf{f}^{\theta}(\mathbf{x})=\mathbf{g}_{l}^{\theta_{l}}(\mathbf{g}^{\theta_{l-1}}_{l-1}(\cdots \mathbf{g}_{1}^{\theta_{1}}(\mathbf{x})))$ to give the output $\mathbf{y}^{\theta}$.

Next, we explain the role of  neural networks in our actor-critic method. We start with the construction of the policies. Although we draw inspiration from recent actor-critic reinforcement learning literature \cite{sac,a3c,PPO}, a deeper description is warranted due to the complexity introduced by the inclusion of memory. The policy $\pi^{\theta^{p}}(x|x',\tau')$ is a probability mass function determining the jumps to new reachable configurations $x$, given that $(x',\tau')$ are the current configuration and the waiting time. The policy is constructed,  by applying a so-called \textit{softmax} function to the outputs of a multi-layered neural network. The softmax function performs a normalizing operation to give a probability mass function. To be concrete, given that the vector $\mathbf{y}^{{\theta}^{p}}$ is the output of an $l$-layered  neural network with input $(x',\tau')$, the $i^{\mathrm{th}}$ component of the softmax function

\begin{equation}
\label{softmax}
  \mathrm{softmax}_{i}(\mathbf{y}^{\theta^{p}}) =  \frac{e^{(y^{\theta^{p}})_{i}}}{\sum_{i} e^{(y^{\theta^{p}})_{i}}},
\end{equation}
gives the corresponding component of the probability mass function $\pi^{\theta^{p}}(x|x',\tau')$, describing the jumps. Here $(y^{\theta^{p}})_{i}$ represents the $i^{\mathrm{th}}$ component of the output vector $\mathbf{y}^{{\theta}^{p}}$. We remark that the construction of an actor with neural-networks followed by a softmax operation is the standard technique for generating discrete stochastic policies in actor-critic reinforcement learning \cite{a3c}.

The structure of $\pi_{x}^{\theta^{q}}(\tau)$ is slightly more complicated as this represents a non-trivial probability density with a positive support. A popular method to construct such complicated densities is by using a mixture-density network \cite{MDN,NN_density_estimation}, where the neural network learns the parameters of a mixture of known distributions. In the present case, the network learns a weighted mixture of gamma densities \cite{Gamma-k,Gamma_mixture2}, which generates the waiting-time distributions. These gamma-distribution mixtures are dense on the positive real line and even a small number of mixture components can be used to create a rich variety of distributions with positive support \cite{Gamma_mixture2}.

In general, a gamma distribution $\gamma_{a,\beta}(\tau)$, with scale parameter $a$ and rate parameter $\beta$ is given by the function
\begin{equation}
\label{gamma_distribution}
\gamma_{a,\beta}(\tau)= \frac{1}{\Gamma(a)} \beta^{a} \tau^{a-1} e^{- \beta \tau},
\end{equation}
where $\Gamma(a)$ is the gamma function calculated at the value $a$. The gamma-mixture density is a weighted sum of such gamma distributions. The weights are normalized and form a probability mass function. The gamma mixture is thus given by
\begin{equation}
 \mathrm{Mixture}(\tau) =    \sum_{i} w_i 
 \gamma_{a_{i},\beta_{i}}(\tau) 
\end{equation}
where the $w_i$ are the normalized weights.

Here the parameters of the gamma mixture are given by a neural network with input $x$.  The network we use branches to give three output vectors $(\mathbf{y}^{\theta^{q}}_{w},\mathbf{y}^{\theta^{q}}_{a},\mathbf{y}^{\theta^{q}}_{\beta})$ representing the weights, scales and rates of the gamma mixture respectively. We finally have
\begin{equation}
\pi^{\theta^q}_{x'} (\tau)   = \sum_{i} \mathrm{softmax}_{i}(\mathbf{y}_{w}^{\theta^{q}}) \gamma_{ (y_{a}^{\theta^{q}})_{i},(y_{\beta}^{\theta^{q}})_{i}}(\tau) ,
\end{equation}
representing the gamma-distribution mixture. The softmax operation is the same as the one described in \eqref{softmax}. The terms $(y_{a}^{\theta^{q}})_{i}$ and $(y_{\beta}^{\theta^{q}})_{i}$ represent the $i^{\mathrm{th}}$ component of the vectors $\mathbf{y}^{\theta^{q}}_{a}$ and $\mathbf{y}^{\theta^{q}}_{\beta}$ respectively.  We thus generate powerful densities to represent waiting-time distributions of the modified dynamics. As an aside, other density-generation methods such as normalizing flows  \cite{Asghar2024,nmf_1,nmf_2} could be used instead. 
 
Our policy structure allows the inclusion of hidden variables, beyond just waiting times. Since the policies distill information from memory to generate decisions, they takes on a role analogous to context networks used in machine learning, especially in large language generation models \cite{context,attention1,attention2}. We seek to explore this connection further in future work.

The value function is approximated using a standard multi-layered feed-forward neural network with weight parameters $\phi$. The inputs to the network are the configuration and the corresponding waiting time. The network outputs an estimation of the value for the inputs given. The actor and critic parameters $\theta^{p}, \theta^{q}, \phi $ are updated using the stochastic gradient descent method, following  Algorithm~\ref{rl_algo}.  In practice, the gradients performed on  neural networks use the modified stochastic gradients given by the ADAM updates \cite{Adam}.

To summarize, we have now constructed a two-policy neural actor-critic framework. The two actors give transition probabilities for the jumps and a probability density function for the waiting time. The neural critic maps the extended space of the process state and waiting time to a value. The relatively simple neural network architectures employ established ideas from widely-used methods \cite{a3c, sac, PPO,MDN} in this new context.

In  neural networks, new information interferes with previously learnt weights in a phenomenon known as \textit{catastrophic forgetting} \cite{catastrophic_Forgetting_review}. This poses a major problem, especially in many reinforcement learning applications where data is obtained sequentially. Constructing neural policies from orthogonal subspaces offers some protection against catastrophic forgetting, which seems to be another advantage of the two-policy structure presented above. Catastrophic forgetting is further mitigated by performing gradient descent over a parallel batch of samples, i.e., \textit{batch learning}. In the present context, selecting a batch size corresponds to selecting the number of sample trajectories for parallel evaluation. Another popular method that could be incorporated in our architecture for reducing interference would be to use a so-called \textit{replay buffer} \cite{experience_replay}. The networks would then trained on randomly chosen samples from the buffer instead of just the immediate previous samples.

Before applying the actor-critic reinforcement learning method in earnest in section 5, we next apply it to the testcase of current fluctuations in the semi-Markov CTRW.

\subsection{Example: semi-Markov CTRW}
\label{semi-Markov_example}
To demonstrate the reinforcement learning method, we consider again the semi-Markov CTRW of section 2.1. The individual waiting-time densities, $\psi^{+}(\tau)$ and $\psi^{-}(\tau)$, of the forward and backward processes in \eqref{ctrw_wtd} are now taken to be gamma distributions  $\gamma_{a,\beta^{+}}(\tau)$ and $\gamma_{a,\beta^{-}}(\tau)$ with the form shown in \eqref{gamma_distribution}. The associated survival functions $\varphi^{+}(\tau)$ and $\varphi^{-}(\tau)$ contain the corresponding upper-incomplete gamma functions. Both distributions are here assumed to have the same value for the scale parameter, $a=2$. In this case, semi-analytical calculations are still possible using an equivalent hidden Markov model as the gamma distribution is an example of a phase-type distribution \cite{Cox_1955_phase_type,phase-type-theory} (see appendix \ref{hmm}). This makes the model an ideal testcase for our algorithm.

%

\begin{figure}
\begin{subfigure}{0.5\textwidth}
\includegraphics[width=0.9\linewidth, height=4cm]{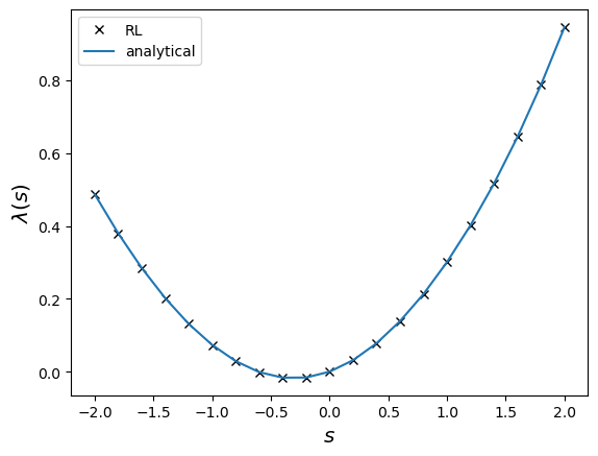}
\caption{}
\label{rw_scgf}
\end{subfigure}
\begin{subfigure}{0.5\textwidth}
\includegraphics[width=0.9\linewidth, height=4cm]{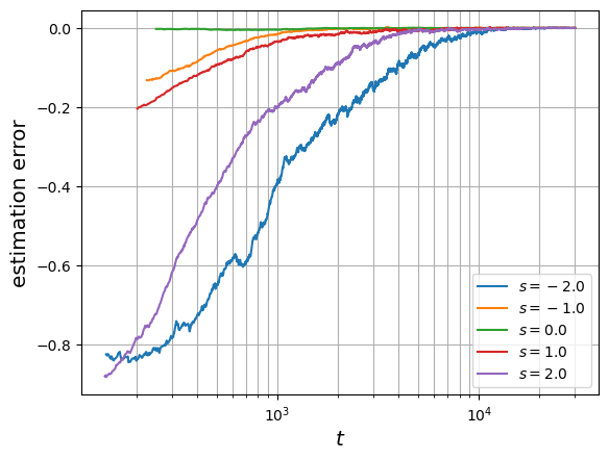}
\caption{}
\label{convergence_errors}
\end{subfigure}
\caption{Current fluctuations in semi-Markov CTRW with $(a,\beta^{+},\beta^{-}) = (2,0.6,0.4)$. (a) Crosses show the estimated SCGF from the actor-critic reinforcement learning algorithm with $t=20\,000$ (and $t_{0}=0$); solid line shows  long-time result from equivalent hidden Markov model.  
(b) Colored lines show convergence of the estimated SCGF to the true asymptotic value, at different values of the conjugate variable $s$, with respect to process time $t$.} 
\label{rw}

\end{figure}

The results in  figure \ref{rw} show excellent agreement of the reinforcement learning with the analytical approach. 
The reinforcement learning also directly gives the modified dynamics.  Figure \ref{rw}(b) shows that the average-reward (the SCGF estimate) converges rapidly to the true value as the process time increases, with slower convergence for values of $s$ corresponding to fluctuations further away from the mean. Preliminary investigations suggest comparable performance with the cloning algorithm of \cite{cavallaro}. The robustness of the reinforcement learning method with respect to key hyperparameters is discussed in appendix \ref{robustness}.

%

\section{Applications}

In this section, we demonstrate the neural reinforcement learning method on more complicated single and multi-particle systems with memory. We particularly focus on semi-Markov analogs of prototypical models that are commonly used as examples to demonstrate far-from-equilibrium physics.

\subsection{Memory-induced ratchets}

 In thermodynamics, ratchet models demonstrate symmetry-breaking mechanisms which may appear surprising and push a system out of equilibrium \cite{Feynman_ratchet,rocking_ratchet, brownianmotors}. A paradigmatic example is the \textit{flashing}  ratchet, see for example \cite{magnasco, flashing_ratchet_review}, where a sawtooth potential is periodically applied to Brownian particles. The particles are trapped in the sawtooth well when the potential is switched on while they diffuse freely when the potential is switched off. In such a system, the asymmetry of the sawtooth potential induces a nonzero current.

A recent area of interest is the examination of ratchet effects in self-propelling systems or active matter. For example, nonzero currents are generated in a ``run-and-tumble" model with symmetric dynamics by applying an asymmetric external potential as in the flashing ratchet \cite{ratchet_run_and_tumble, Angelani_2011}. Here, we introduce another type of ratchet which does not need an external potential. Instead, memory creates asymmetry in the dynamics and a nonzero current is generated.

We consider here the case where a particle performs a CTRW on a one-dimensional periodic lattice with random flips in its orientation. This is reminiscent of run-and-tumble dynamics, but with a semi-Markov complication. Since run-and-tumble systems are archetypal models used for active matter \cite{chemotaxis2,chemotaxis1, run_and_tumble_nature}, inclusion of memory could help in understanding biological transport such as the motility of bacteria. Indeed, as pointed out in \cite{non_markov_run_and_tumble2} one might anticipate that memory plays at least some role in realistic environments; see also \cite{Generalize_run_and_tumble} for further discussion on generalized run-and-tumble processes and applications.

\begin{figure}
\centering
\includegraphics[scale=0.8]{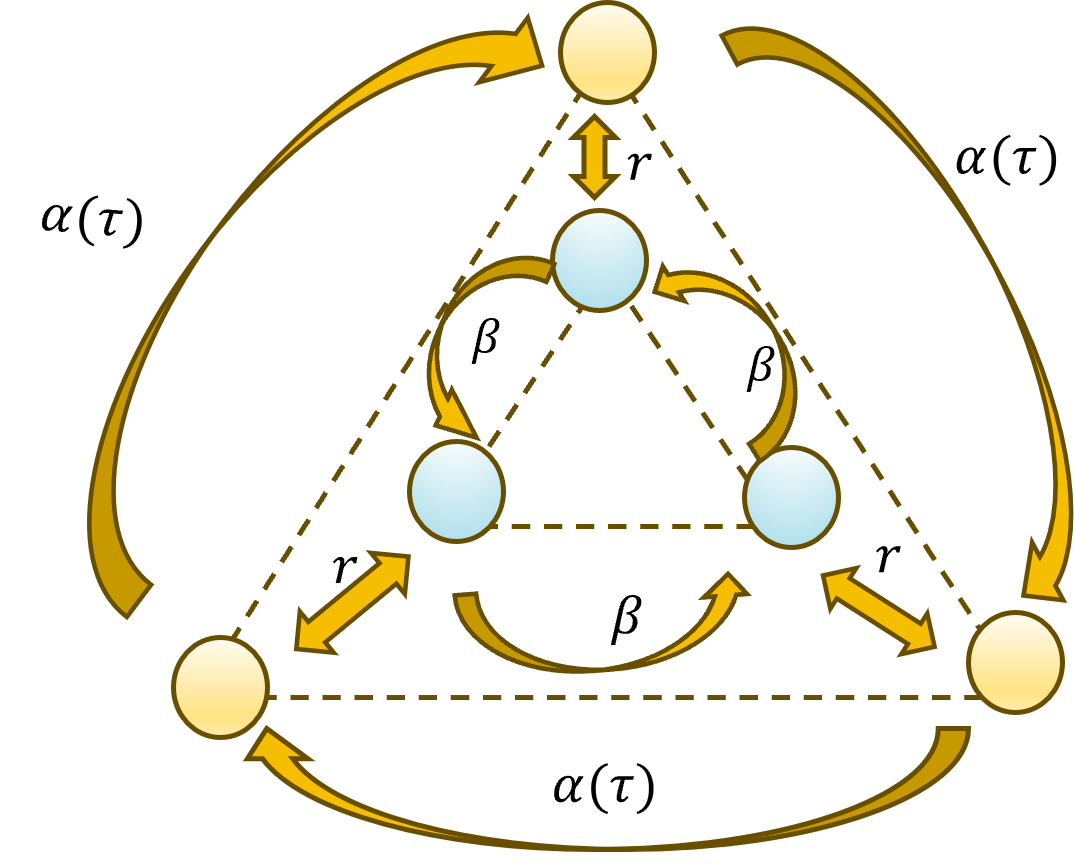}
\caption{Representation of the memory-induced ratchet on a three-site ring. The outer ring shows the forward semi-Markov transitions. The inner ring shows the backward Markov transitions. The particle switches between the rings with constant transition rate $r$. }
\label{ratchet_rep}
\end{figure}

To be concrete, we consider a particle which is constrained to move in a particular direction (a run) until a shift in directional polarity (a tumble) occurs. The system is easily visualized by the three site ring in figure \ref{ratchet_rep}. 
As in the previous CTRW case, since all the sites are identical, the dynamics of the current (net number of clockwise jumps) does not depend on the number of sites in the ring. In figure \ref{ratchet_rep}, the outer and inner rings show the forward and backward orientated dynamics respectively. A change in orientation is represented by a jump between the rings.  

The system is controlled by three stochastic clocks associated with the forward jump, backward jump and the next directional change respectively. If all three clocks are exponentially distributed then a nonzero current is only possible when the means of the forward and backward distributions are different. However,  even if the two distributions have the same mean, it can be shown that nonzero currents may be generated as long as the forward and backward clocks follow different distributions. This is the ratchet effect. We save the detailed examination of this effect for future work; here we concentrate on computing the current fluctuations.

As a minimal model we take just one of the clocks as nonexponential. Without loss of generality, we assume that the forward clock corresponds to a time-varying rate $\alpha(\tau)=\psi^{\alpha}(\tau)/\varphi^{\alpha}(\tau)$, where $\psi^{\alpha}(\tau)$ is a nonexponential distribution and $\varphi^{\alpha}(\tau)$ is the associated survival function. The distribution of the clocks triggering the backward jump and the reorientations are exponential with constant rates $\beta$ and $r$ respectively.\footnote{In the run-and-tumble language, we thus have memory in the run dynamics but not in the tumble statistics, in contrast to the model of \cite{non_markov_run_and_tumble2}.}  We stress that the system moves only in one direction until a reorientation event occurs, differentiating the model from the bidirectional CTRW of section 4.2. The waiting times of the system during the forward and backward phases are respectively given as

\begin{equation}
\psi_{x^{+}}(\tau)= \psi^{\alpha}(\tau)e^{-r\tau}+re^{-r\tau} \varphi^{\alpha}(\tau), \quad \psi_{x^{-}}(\tau)= (r+\beta)e^{-(r+\beta)\tau}.
\end{equation}

\begin{figure}
\begin{subfigure}{0.5\textwidth}
\includegraphics[width=0.9\linewidth, height=4cm]{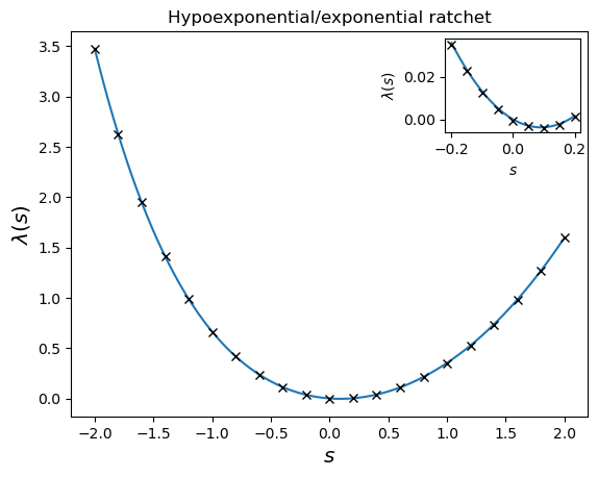}
\caption{}
\label{ratchet_hypo}
\end{subfigure}
\begin{subfigure}{0.5\textwidth}
\includegraphics[width=0.9\linewidth, height=4cm]{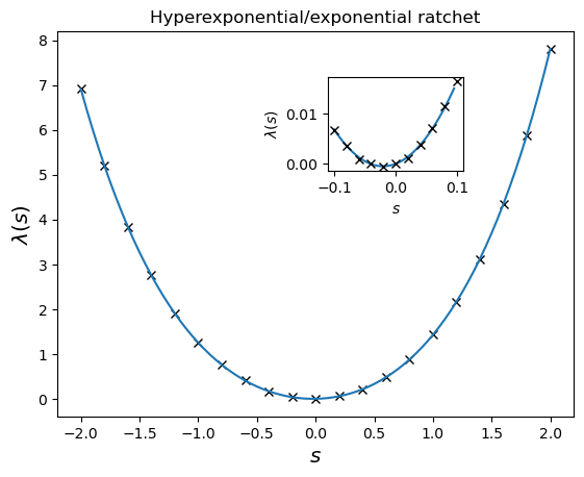}
\caption{}
\label{hyper_ratchet}
\end{subfigure}
\caption{ SCGF for currents in memory-induced ratchet systems with nonexponential distribution for forward clock and exponential distribution with same mean for backward clock. Crosses show results from reinforcement learning ($t=20\,000$, $t_0=0$); lines show predictions from equivalent hidden Markov models. (a) Hypoexponential distribution for forward clock, with parameters $(\alpha_{1}  ,\alpha_{2})=(1,2)$; exponential distribution for reorientation with rate $r=1$. (b)  Hyperexponential distribution for forward clock where density is a weighted sum of two equiprobable exponential distributions with rates $(\alpha_{1},\alpha_{2})=(1,2)$; exponential distribution for reorientation with rate $r=2$.}
\label{ratchet_results}
\end{figure}

The computational framework of the previous section was applied to obtain the SCGF of the currents in this system for certain choices of phase-type distributions for the forward clock. The results are shown in figure \ref{ratchet_results}. The plots show good agreement with the analytical approach via equivalent hidden Markov models (see appendix \ref{hmm}). The SCGF also indirectly shows the nonzero mean current in the system and its direction. Specifically, the mean current is given by the slope of the SCGF computed at the point where the conjugate variable $s$ is zero.  Choosing the forward distribution as a hypoexponential gives a negative average current as shown by inset in figure \ref{ratchet_results}(a). Alternatively, when a hyperexponential distribution is chosen, the inset plot of figure \ref{ratchet_results}(b) shows that  a positive current is observed. 

We emphasize that the nonzero mean currents are a direct result of the memory in the system; if the jumps in forward and backward runs both had exponential waiting times with the same mean, then there would be no current in the system regardless of the tumbling rate. Beyond the mean, the effect of memory on the fluctuations is also seen here in the asymmetry of the SCGF. In particular, the Gallavotti-Cohen fluctuation relation \cite{GC_symmetry} does not hold, i.e., there is no constant $\epsilon$ for which $\lambda(s)=\lambda(\epsilon-s)$.

\subsection{Memory-dependent totally asymmetric exclusion processes}

In this section, we move beyond single-particle dynamics and show how to apply the reinforcement learning framework to multi-particle systems. In particular, we choose the well-known asymmetric exclusion process (ASEP) \cite{ASEP} for our demonstrations. As a one-dimensional interacting particle system, the exclusion process (and its variants) has been commonly used as a testing ground for studying nonequilibrium systems. For an overview of theoretical work and connections to real-world applications see \cite{asep_review_1,asep_review2} and references therein. In particular, exclusion processes with hidden states (or, correspondingly, nonexponential waiting times) are important for modeling both ribosome translation and transport of biological motors   \cite{exclusion_non_markov,exclusion_process_non_markov2,Gorrison_rna}.

\begin{figure}
\centering
\includegraphics[scale=0.75]{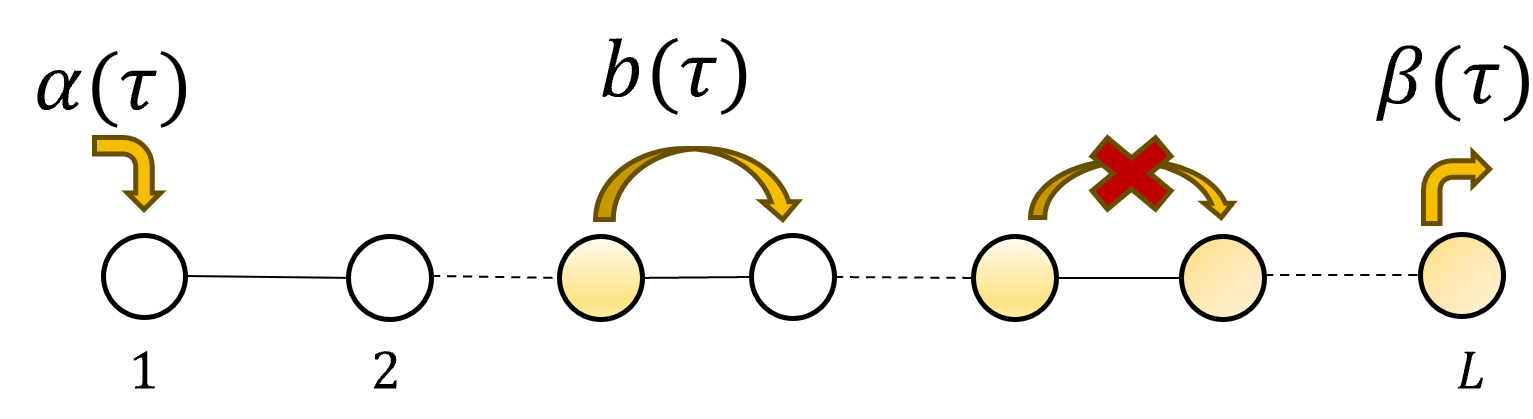}
\
\caption{Memory-dependent TASEP with particles indicated by colored circles. In general, the arrival, bulk and departure processes are non-Markovian with time-varying rates. Arrows encode the exclusion rules; a particle is only allowed to enter a site when it is unoccupied.}
\label{tasep_rep}

\end{figure}

We now describe more precisely the process of interest. At any given time, each site on the one-dimensional lattice site is allowed to be occupied by at most one particle; particles can hop to the neighboring sites only if they are vacant. When the particle dynamics is unidirectional, the system is known as the \textit{totally asymmetric} exclusion process (TASEP). We will be particularly interested in the open-boundary version of the model where it has been shown that the bond currents even in the Markov case display rich statistical behavior \cite{ASEP_current}. Here we look at currents in memory-dependent TASEPs in the vein of \cite{cavallaro,Harris_2015,Gorrison_rna}. Figure \ref{tasep_rep} shows such a system. 

The dynamics of an $L$-site TASEP is described by transitions in the state space $\mathcal{S}$ where each configuration is represented by a vector $\boldsymbol{\eta}$ with $L$ elements. When a lattice site is occupied, the corresponding component of the configuration vector $\boldsymbol{\eta}$ takes the value one, while it is zero when the site is vacant. We introduce memory dependence by considering that the waiting times at each configuration follow a nonexponential distribution.  We denote the nonexponential distributions giving the waiting times of arrivals, bulk transitions and departures by $\psi^{\alpha}(\tau)$, $\psi^{b}(\tau)$  and $\psi^{\beta}{(\tau)}$ respectively. The corresponding survival functions $\varphi^{\alpha}(\tau)$, $\varphi^{\beta}(\tau)$ and $\varphi^{b}(\tau)$ give the individual rates

\begin{equation}
 \alpha(\tau) = \frac{\psi^{\alpha}(\tau)}{\varphi^{\alpha}(\tau)}, \quad \beta(\tau)=\frac{\psi^{\beta}(\tau)}{\varphi^{\beta}(\tau)}, \quad
 \ b(\tau)= \frac{\psi^{b}(\tau)}{\varphi^{b}(\tau)}.
  \end{equation}

 A useful method to track this system is to again imagine stochastic clocks that trigger the corresponding transitions. In the non-Markov TASEP, it is important to distinguish between attaching clocks to the site and attaching the clocks to the particles, as the choice may lead to different dynamics. 
We consider that a clock is only active when its associated transition is allowed. Explicitly, the exclusion rules give the following conditions: (i) the clock for the arrivals is only active when the first site is empty, (ii) a clock for a transition out of a site in the bulk is only active when that site is occupied and the next site is vacant  (iii) the clock for the departures is active as long as the last site is occupied. When a transition associated with a clock occurs, the clock becomes inactive. When the transition is next allowed, the clock is reset and becomes active again. 

 Given that $\tau^{0}$, $\tau^{L}$ and $\tau^{i}$ (for $ i \in \{1,2, \dots,L-1\}$) are the times since the arrival, departure and bulk-transition clocks have been active, the combined waiting-time distribution at a configuration $\boldsymbol{\eta}$ of the memory-dependent TASEP is  

\begin{align}
\label{multi_tasep}
\psi_{\boldsymbol{\eta}}(&\tau |\tau^{0},\tau^{1},...,\tau^{L})  \nonumber \\ = 
& \left( (1-\eta_1) \frac{\psi^{\alpha}(\tau+\tau^0)}{\varphi^{\alpha}(\tau+\tau^{0})} + \sum_{i=1}^{L-1} \eta_i (1-\eta_{i+1}) \frac{\psi^{b}(\tau +\tau^{i})}{\varphi^{b}(\tau+\tau^{i})} + \eta_{L}\frac{\psi^{\beta}(\tau +\tau^{L})}{\varphi_{\beta}(\tau+\tau^{L})} \right)
 \nonumber \\  & \times \mathrm{exp} \left( (1-\eta_1) \ln\frac{\varphi^{\alpha}(\tau+\tau^0)}{\varphi^{\alpha}(\tau^{0})} + \sum_{i=1}^{L-1} \eta_i (1-\eta_{i+1}) \ln \frac{\varphi^{b}(\tau+\tau_i)}{\varphi^{b}(\tau^{i})} \right. \nonumber \\& \phantom{0000000000} \left. + \eta_{L} \ln\frac{\varphi^{\beta}(\tau+\tau^{L})}{\varphi^{\beta}{(\tau^{L})}} \right).
\end{align}
Here $\eta_{i}$ (for $i \in \{1,2,\cdots,L\}$) are the elements of the vector $\boldsymbol{\eta}$.

The configuration space of multi-particle systems generally scales exponentially with the number of sites. For example, in the open-boundary $L$-site exclusion process discussed above, the  state space contains $2^{L}$ configurations, leading to a drop in efficiency of optimization methods due to the well-known \textit{curse of dimensionality} \cite{bellman_dynamic_programming}. Neural networks have allowed the application of reinforcement learning  in large systems  \cite{dqn,alphago,a3c, PPO} by adopting specialized neural units such as recurrent  neural networks (RNNs) and convolution  neural networks (CNNs) commonly used in other machine learning domains including image, speech and natural language processing \cite{image, audio_speech,BERT_language, attention1, NN}. We will use networks based on the so-called gated recurrent unit (GRU), an RNN variant, for handling large state spaces in the memory-dependent TASEP of section 5.2.2 (see Appendix \ref{RNN method} for more details).

\subsubsection{Two-site TASEP with gamma-distributed waiting times}

We first showcase the reinforcement learning method for a minimal exclusion process on just two sites. The system dynamics is governed by three memory-dependent transitions: the arrival into the first site, a (bulk) transition from the first site into the second, and the departure from the second site.  We choose gamma-distributed waiting times (with scale parameter $a=2$) for each transition, motivated by the biological system in \cite{Gorrison_rna} where gamma distributions have been used to model ribosome dwell times.

The combined waiting-time distribution, is then obtained from \eqref{multi_tasep} with $i=2$ and the relevant gamma distributions for the individual transitions. Focusing now on the arrival currents (by weighting jumps into the system by $e^{s}$), we computed the SCGF with the reinforcement learning algorithm.  Note that to apply the reinforcement learning method, the extended state space must also include the times since each of the arrival, bulk and departure clocks have been active before the process entered the present configuration. In practice, we track the state space $(\boldsymbol{\eta},\tau+\tau^0,\tau+\tau^{1}, \tau+\tau^{2})$, with implicit dependence on $\tau$, to apply Algorithm~\ref{rl_algo}.

\begin{figure}
\centering
\includegraphics[scale=0.5]{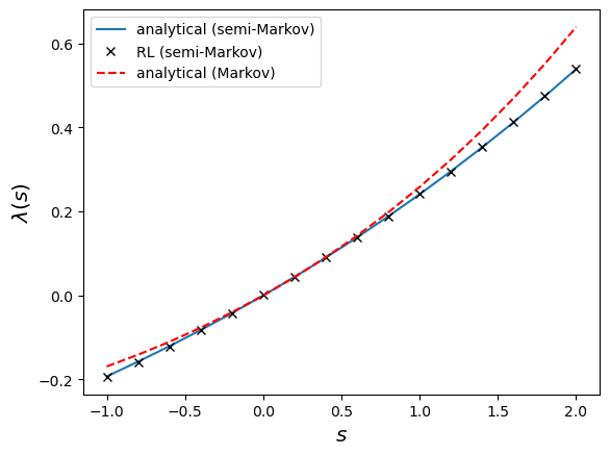}
\caption{SCGF of arrival currents in two-site gamma TASEP, with $(a,\alpha,b,\beta)= (2,2,1,0.8)$. Here $a$ is the scale parameter for all the gamma waiting-time distributions, while $\alpha$, $b$ and $\beta$ are separate rate parameters for the arrival departure and bulk transitions respectively. Crosses are results from the reinforcement learning method ($t_0=0$, $t=50\,000$) and solid line is the prediction from the equivalent hidden Markov model. The dotted line shows the SCGF of arrival currents in a Markovian two-site TASEP where each transition has an exponential waiting time with the same mean as the corresponding transition in the gamma TASEP. }
\label{tasep_gamma}
\end{figure}

The results in  figure \ref{tasep_gamma} again show excellent agreement with the analytical approach based on hidden Markov models (see appendix \ref{hmm}). For comparison, in figure \ref{tasep_gamma} we also display the SCGF for a Markovian TASEP with exponentially-distributed waiting times where each transition has the same mean as the corresponding gamma-distributed transition of the semi-Markov model. We clearly see the effect of the memory on rare events; fluctuations far away from the mean are quite different in the two cases even though those close to the mean (with small $s$) are similar.

\subsubsection{Many-site TASEP with gamma-distributed arrival times}
\label{many-site TASEP}

We conclude this section with a preliminary investigation of the neural reinforcement learning method for large systems, by extending the semi-Markov TASEP beyond two sites. To simplify calculations, we now consider that only the waiting times of the arrivals into the system are nonexponential and that all other transitions remain Markovian. We again choose a gamma distribution, with scale parameter $a=2$, to model the waiting times of the arrivals into the system. Choosing Markovian rates for bulk and  departure transitions as $b$ and $\beta$ respectively, the expression \eqref{multi_tasep} giving the combined waiting time at each configuration of the process reduces to   
\begin{align}
\label{multi_tasep_2}
\psi_{\boldsymbol{\eta}}(&\tau |\tau^{0})  \nonumber \\ = 
&\left( (1-\eta_1) \frac{\psi^{\alpha}(\tau+\tau^0)}{\varphi^{\alpha}(\tau+\tau^{0})} + \sum_{i=1}^{L-1} \eta_i (1-\eta_{i+1}) b + \eta_{L} \beta \right)
 \nonumber \\  & \times \mathrm{exp} \left( (1-\eta_1) \ln\frac{\varphi^{\alpha}(\tau+\tau^0)}{\varphi^{\alpha}(\tau^{0})} - \left ( \sum_{i=1}^{L-1} \eta_i (1-\eta_{i+1}) b + \eta_{L} \beta \right) \tau \right),
\end{align}
where $\psi^{\alpha}(\tau +\tau^{0})$ and $\varphi^{\alpha}(\tau+\tau^{0})$ now describe a gamma distribution and its corresponding survival function. As in the previous example, counting the fluctuations of currents into the system involves weighting of the arrivals by the factor $e^{s}$.

\begin{figure}
\centering
\includegraphics[scale=0.75]{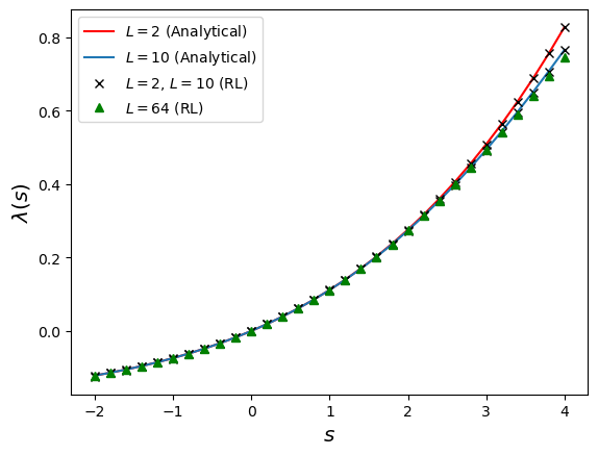}
\
\caption{SCGF of arrival currents in TASEP with gamma-distributed arrival times for system sizes, $L=2$, $L=10$ and $L=64$. All cases have parameters $(a,\alpha,b,\beta) =(2,0.2,1,1)$ and simulation times $(t_0,t)=(0,150 
  \,000)$. Crosses show results computed with reinforcement learning for $L=2$ and $L=10$, with blue and red lines showing the corresponding SCGF values calculated via exact diagonalization. Green triangles show reinforcement learning results for $L=64$ where exact diagonalization is not possible.}
\label{tasep_gamma_multisite}

\end{figure}

Figure \ref{tasep_gamma_multisite} shows the SCGF computed for different system sizes: $L=2$, $L=10$ and $L=64$. For $L=2$ and $L=10$, the theoretical results for comparison were obtained via an analytical approach involving the exact diagonalization of equivalent hidden-Markov models, similar to the procedure in appendix \ref{hmm}, but with the application of tensor products to simplify calculations for $L=10$. 
The SCGF calculations using the exact diagonalization method become difficult beyond $L=10$. However, significantly, the neural reinforcement learning still produces results for $L=64$ demonstrating the potential of this method for application to large systems, much beyond the scope of exact diagonalization.

Next we briefly comment on the physics of the system to interpret the SCGF in figure \ref{tasep_gamma_multisite} and to justify that the $L=64$ results are reasonable. First note that we deliberately choose the parameters of the gamma distribution such that the rate of arrivals is much smaller than the bulk and the departure rates. This means that the stationary occupancy of the first site is low and the mean arrival current is expected to be asymptotically independent of the system size, just as in the low-density regime of the corresponding Markovian model \cite{ASEP_current}. The modified dynamics for nonzero $s$ effectively involves altering the rate of arrivals due to the weighting $e^{s}$. For small-magnitude $s$ (fluctuations close to the mean), figure \ref{tasep_gamma_multisite} shows that the curves for $L=2$, $L=10$ and $L=64$ are indeed essentially indistinguishable. The same is true for large negative $s$, corresponding to reducing the effective rate of arrivals. On the other hand, for large positive values of $s$, the effective arrival rates become comparable to the other rates and the SCGF becomes more strongly dependent on the system size. Nevertheless, just as in the Markovian model we expect convergence to an $L$-independent limiting form. The close agreement of $L=64$ and $L=10$ in the numerical results is consistent with this intuition. Furthermore, the SCGF appears to become linear for large values of $s$ (with slope approximately $b/4$) which is indicative of a dynamical phase transition to a maximal-current phase, as seen in the Markovian case. Of course, these results are not surprising considering that only the arrivals are memory-dependent. However, they do evidence the reliability of neural reinforcement learning for this type of large system.

The reinforcement learning was implemented on the extended state space $(\boldsymbol{\eta},\tau, \tau^{0})$: the lattice configuration, the waiting time at the configuration and a memory variable indicating the time since the arrival clock has been active before entering the current configuration. The optimal transitions $(\boldsymbol{\eta'},\tau',\tau'^{0}) \to (\boldsymbol{\eta},\tau,\tau^{0})$ can be learnt by policies of form $\pi^{\theta^{p}} (\boldsymbol{\eta}|\boldsymbol{\eta'},\tau',\tau^{0})$ and $\pi_{\boldsymbol{\eta},\tau^{0}}^{\theta^{q}}(\tau)$. The first policy updates the configuration and the arrival clock memory while the second policy generates the waiting time in the new configuration. With this policy structure, Algorithm~\ref{rl_algo} of section \ref{diff_act_crit} can be directly applied on the extended state space. For this more complicated model, the neural reinforcement learning method was carried out using recurrent neural networks, details of which are presented in appendix \ref{RNN method}.

\section{Summary and outlook}

In this work, we demonstrated an application of neural reinforcement learning for rare-event computations in memory-dependent systems. We focused on current fluctuations in  continuous-time jump processes with nonexponential waiting-time distributions. 

In particular, we extended the reinforcement learning framework given in \cite{rose} and showed how artificial  neural networks can be used as powerful representational frameworks for constructing policies. Viewing the reinforcement learning problem as a sequential optimization task on the extended state space of configurations and waiting times, we considered a two-policy structure. Apart from helping protect the networks from catastrophic forgetting, this two-policy structure also allows the inclusion of hidden variables. 

We demonstrated results for a variety of memory-dependent models including single-particle systems with ratchet effects and many-particle interacting systems. The models chosen were rare examples of non-Markovian systems where analytical results for the SCGF are accessible and the reinforcement learning outputs showed convincing agreement.

This suggests that the method could be a powerful tool in the more common non-Markov situations where analytical results are not available. The two-policy structure is expected to be particularly crucial for waiting-time distributions that are strongly nonexponential. Note, the present method is limited to processes with time-homogeneous dynamics (at least in the long-time limit) and a well-defined stationarity in the extended state space containing the hidden variables. The modification of the method for non-Markov processes which are not stationary in the extended state space, such as some models with current dependence \cite{ERW_shutz,Kenkre_ERW}, is a topic for future work.

An immediate next step towards the integration of machine learning methods in nonequilibrium analysis is to undertake a formal benchmarking study, comparing amongst neural network architectures and against existing algorithms such as cloning and transition path sampling. Development of benchmarking metrics might even allow for combination of the reinforcement learning framework with algorithms like cloning in order to obtain optimal performance. An important area of focus for performance improvements is rare-event analysis in large systems. An important aspect here is to make improvements to the performance efficiency of deep learning architectures such as the RNN-based actor-critic shown in appendix \ref{RNN method}. Recently, representational frameworks known as tensor networks, borrowed from quantum physics, have gained popularity for efficient processing of nonequilibrium systems with large state spaces \cite{Tensor_Networks_Garrahan,  gillman2022reinforcement}. A hybrid neural network and tensor network architecture in the vein of \cite{tensor_NN} can be envisioned where the policy controlling the jumps is replaced with a tensor network.

In addition to computational advances, machine learning presents a promising approach for investigating the physics of nonequilibrium processes. One area of topical research is the properties of dynamical phase transitions both in Markov and non-Markov systems \cite{phase_transitions,Gorrisen_DMRG, Garrahan_phase_transitions}. Some initial work with Markovian zero-range process suggests that the neural reinforcement learning method may be used in the detection of phase transitions in the non-Markov case as well. This is an area for future study, possibly in the context of the many-site TASEP with nonexponential arrival times, as in section \ref{many-site TASEP}.     

Another interesting direction to pursue is the extension of the computational framework beyond the method presented in this paper to study rare events in non-Markov systems where the large-deviation principle \eqref{LDP} has a different speed. This is related to the earlier discussion of time-inhomogeneous dynamics, indeed some current-dependent models, such as variants of the elephant random walk, exhibit non-standard large-deviation principles \cite{Harris_erw, jack_harris}. The few theoretical and numerical results for current fluctuations in such scenarios are available, which could perhaps be tested and generalized by the application of advanced machine learning methods.

In conclusion, we believe that our study helps demonstrate the potential of deep reinforcement learning techniques for analysis of nonequilibrium processes and thus opens the door for further exploration of non-Markov systems; from a statistical physics viewpoint, the SCGF crucially reveals the effect of memory on rare events.

\appendix

\section{SCGF from hidden Markov models}
\label{hmm}

We show here an analytical approach for calculating the SCGF of the current fluctuations in the semi-Markov models discussed in the main text.  The waiting times here follow simple phase-type distributions such as the gamma, hypoexponential and hyperexponential distributions which means that semi-analytical calculation of the SCGF is possible from the equivalent hidden Markov description. Models with such waiting-time distributions can be considered as \textit{course-grained} representations of Markov systems with hidden sites and/or transitions. The current fluctuations can then  be obtained by computing the principal eigenvalue of the Markov generator.

\subsection{Background on phase-type distributions}

\begin{figure}
\begin{subfigure}{0.5\textwidth}
\centering
\includegraphics[width=0.5\linewidth, height=1.0cm]{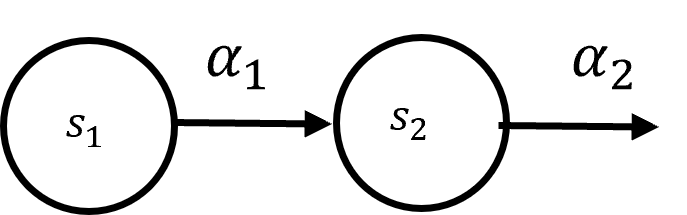}
\caption{}
\label{series}
\end{subfigure}
\begin{subfigure}{0.5\textwidth}
\centering
\includegraphics[width=0.5\linewidth, height=2.5cm]{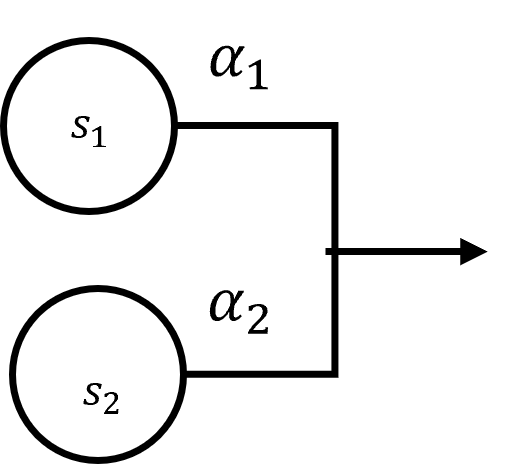}
\caption{}
\label{parallel}
\end{subfigure}
\caption{Phase-type distributions. (a) Exponential distributions connected in series to generate a hypoexponential ($\alpha_1 \neq \alpha_{2})$ or a gamma ($\alpha_1= \alpha_2$) distribution; {$s_2$ is a hidden state not seen at the course-grained level}. (b) Exponential distributions connected in parallel to generate a hyperexponential distribution; the branch from $s_1$ is chosen with weight $w_1$ and the branch from $s_2$ is chosen with weight $w_2$, but the two separate branches are hidden at the course-grained level. }
\label{connection}
\end{figure}

Phase-type distributions are, by definition, obtained by convolution or mixture of exponential distributions \cite{Cox_1955_phase_type}. For example, a gamma density
\begin{equation}
\gamma_{a,\alpha}(\tau)= \frac{ \alpha^{a} \tau^{a-1}}{\Gamma(a)} e^{-\alpha \tau}
\end{equation}
is formed by the convolution of $a$ exponential distributions with the same rate $\alpha$. For $a=2$ we have

\begin{equation}
\gamma_{2,\alpha}(\tau) = \int_{0}^{\tau} \alpha^{2} e^{-\alpha u} e^{-\alpha (\tau-u)} du,
\end{equation}
which is the coarse-graining of two exponential distributions connected in \textit{series} as shown in figure \ref{connection}(a). Choosing different rates for the exponential distributions gives the generalized gamma or the \textit{hypo}exponential distribution
\begin{align}
\mathrm{Hypo}_{2,\alpha_1,\alpha_2}(\tau) & =  \int_{0}^{\tau} \alpha_{1}\alpha_{2} e^{-\alpha_{1} u} e^{-\alpha_{2} (\tau-u)}du \nonumber \\  & =\frac{\alpha_{1} \alpha_{2}}{\alpha_1-\alpha_2} \left(e^{\alpha_{1}\tau}-e^{-\alpha_{2}\tau}\right), \quad  \alpha_{1} > \alpha_{2}.
\end{align}

The coarse-graining of exponential distributions connected in \textit{parallel} as in figure \ref{connection}(b) leads to another phase-type distribution known as the \textit{hyper}exponential distribution. The parallel combination is expressed  as the weighted sum or mixture of exponential distributions

\begin{align}
\mathrm{Hyper}_{2,\alpha_1,\alpha_2}(\tau)  =  w_{1} \alpha_{1} e^{-\alpha_{1} \tau} + w_{2} \alpha_2 e^{-\alpha_{2}\tau},  
\end{align}
where $w_1$ and $w_2$ are normalized weights giving the probability of choosing a branch.

\subsection{Semi-Markov CTRW}

\begin{figure}
\centering
\includegraphics[scale=0.75]{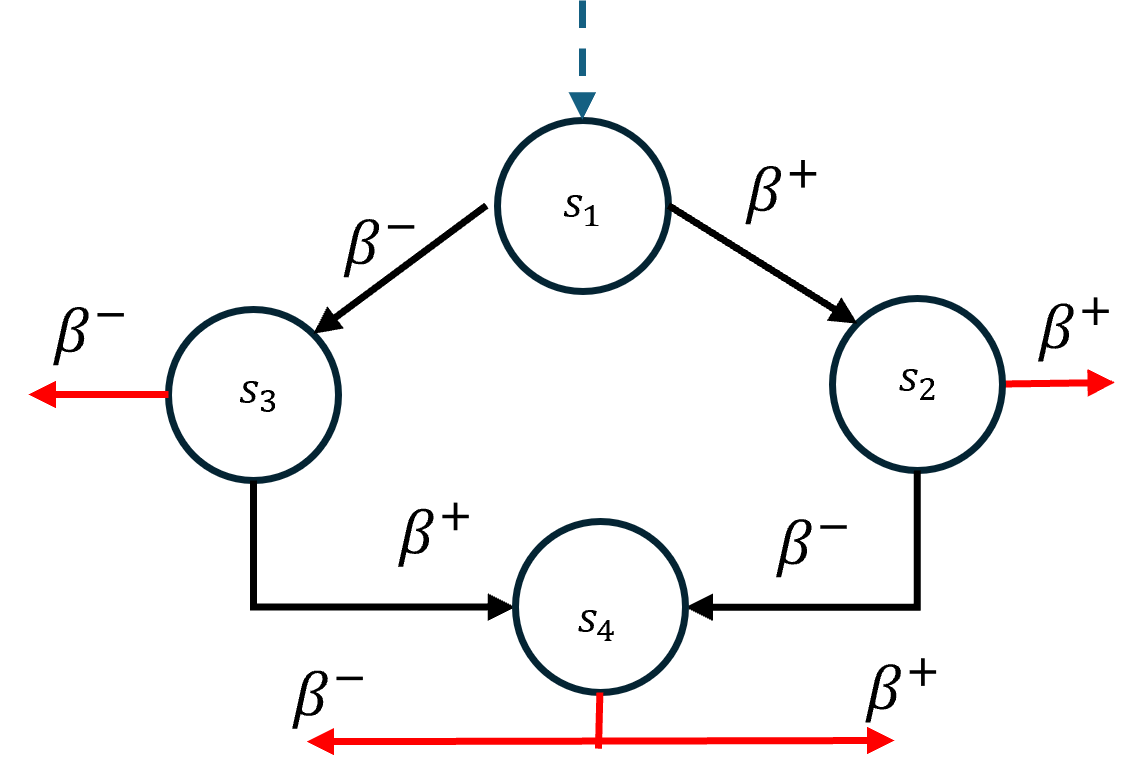}
\caption{A hidden Markov representation of a cell in a semi-Markov CTRW. The particle exit is modeled in three Markov stages starting from $s_1$ and moving down to $s_4$. The red arrows indicate outputs from the cell. The dotted arrow indicates input from the previous cell. In the periodic case, the cell output is looped back into the initial state $s_1$.}
\label{semi_markov_rw_hmm}
\end{figure}

In this subsection, we show in detail how the hidden Markov structure can be used to calculate current fluctuations in the periodic semi-Markov random walk with gamma-distributed waiting times. To simplify analysis, all the gamma distributions were assumed to have the same scale parameter, $a=2$. The two competing gamma distributions triggering the forward and backward transitions at each lattice site are modeled over the four-state cell structure shown in figure~\ref{semi_markov_rw_hmm}. In other words, each course-grained  lattice site is replaced by a cell with four internal states.  
The system lattice can now be visualized as a ring of such cells. A similar cell structure was used in \cite{maes_hidden_semi_Markov_fluctuations}. 

A particle exits a lattice cell via a three-stage Markov process starting in state $s_{1}$.   In the first stage, a move to either $s_{2}$ or $s_{3}$ is triggered by competing exponential distributions with rates $\beta^{+}$ and $\beta^{-}$ respectively. In the second stage, the particle immediately exits the cell if the  distribution that triggered the transition in the first stage wins again. Otherwise, the particle moves to state $s_4$. In the third stage, a forward or backward exit occurs from state $s_4$, with rates $\beta^{+}$ and $\beta^{-}$ respectively. After the exit, depending on whether a forward or backward jump occurs, the particle then enters state $s_1$ of the next or preceding cell in the ring. For a three-site case, the dynamics is then given by a $12 \times 12$ generator matrix.  However, it is possible to reduce the matrix, since we are only interested in the statistics of the system current. As the transition probabilities are independent of the position on the ring, the current dynamics is homogeneous with respect to the cells. We may assume all exits from a cell loop back to state $s_1$ of the same cell and consider a simple $4 \times 4$ Markov generator matrix
\begin{equation}
G = \begin{bmatrix} -\beta^{+}-\beta^{-} & \beta^{+} & \beta^{-} & 0  \\  \beta^{+} & -\beta^{+}-\beta^{-} & 0 & \beta^{-} \\ \beta^{-} & 0   & -\beta^{+}-\beta^{-}  & \beta^{+} \\
\beta^{+}+\beta^{-} & 0 & 0  & -\beta^{+}-\beta^{-} 
\end{bmatrix}.
\end{equation}

We obtain the SCGF for the current fluctuations of the above hidden Markov process following the standard procedure given for example in \cite{touchette2}. Weighting forward and backward transitions out of the cell (shown with red arrows in figure \ref{semi_markov_rw_hmm}) with the exponential factors $e^{s}$ and $e^{-s}$, gives the tilted generator matrix    

\begin{equation}
\tilde{G}(s) = \begin{bmatrix} -\beta^{+}-\beta^{-} & \beta^{+}  & \beta^{-} & 0  \\  \beta^{+} e^{s} & -\beta^{+}-\beta^{-} & 0 & \beta^{-}  \\ \beta^{-} e^{-s} & 0   & -\beta^{+}-\beta^{-}  & \beta^{+} \\
\beta^{+} e^{s}+\beta^{-} e^{-s} & 0 & 0  & -\beta^{+}-\beta^{-} 
\end{bmatrix}.
\end{equation}
Finally, the principal eigenvalue of the above generator matrix gives the required SCGF. We obtain this numerically for comparison against the reinforcement learning results in the main text.

\subsection{Memory-induced ratchets}

Semi-analytical results for current fluctuations in the hypoexponential and hyperexponential ratchets of section 5.1 were calculated by following a procedure similar to the one shown in the previous section.

\begin{figure}
\begin{subfigure}{0.5\textwidth}
\includegraphics[width=0.9\linewidth, height=3cm]{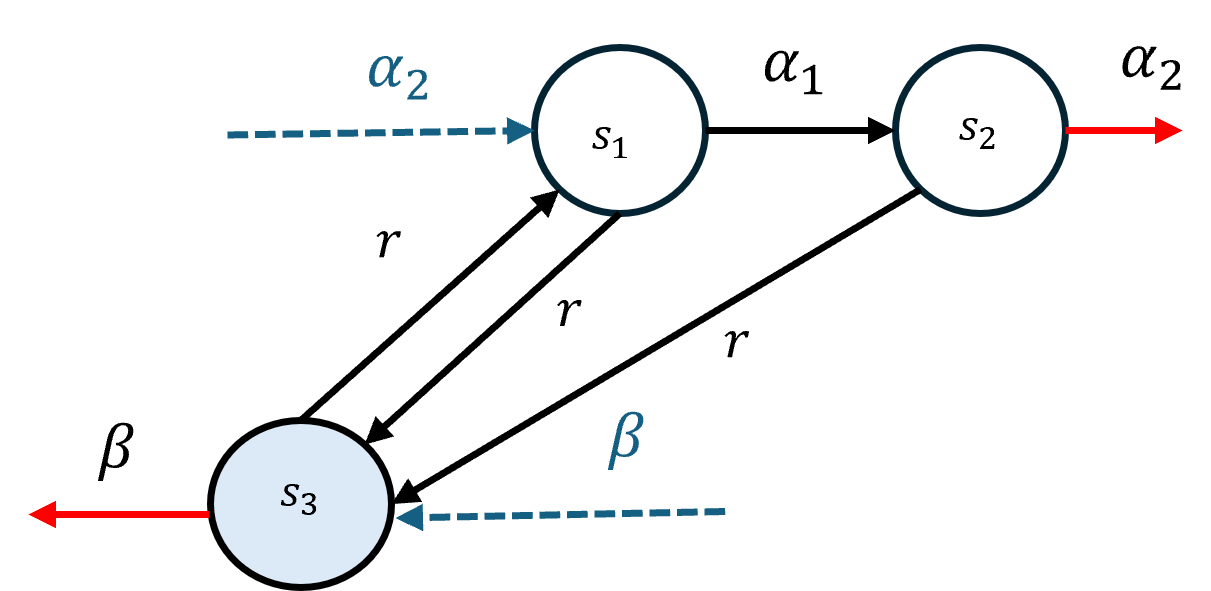}
\caption{}
\label{hypo_ratchet_hmm}
\end{subfigure}
\begin{subfigure}{0.5\textwidth}
\includegraphics[width=0.9\linewidth, height=3.5cm]{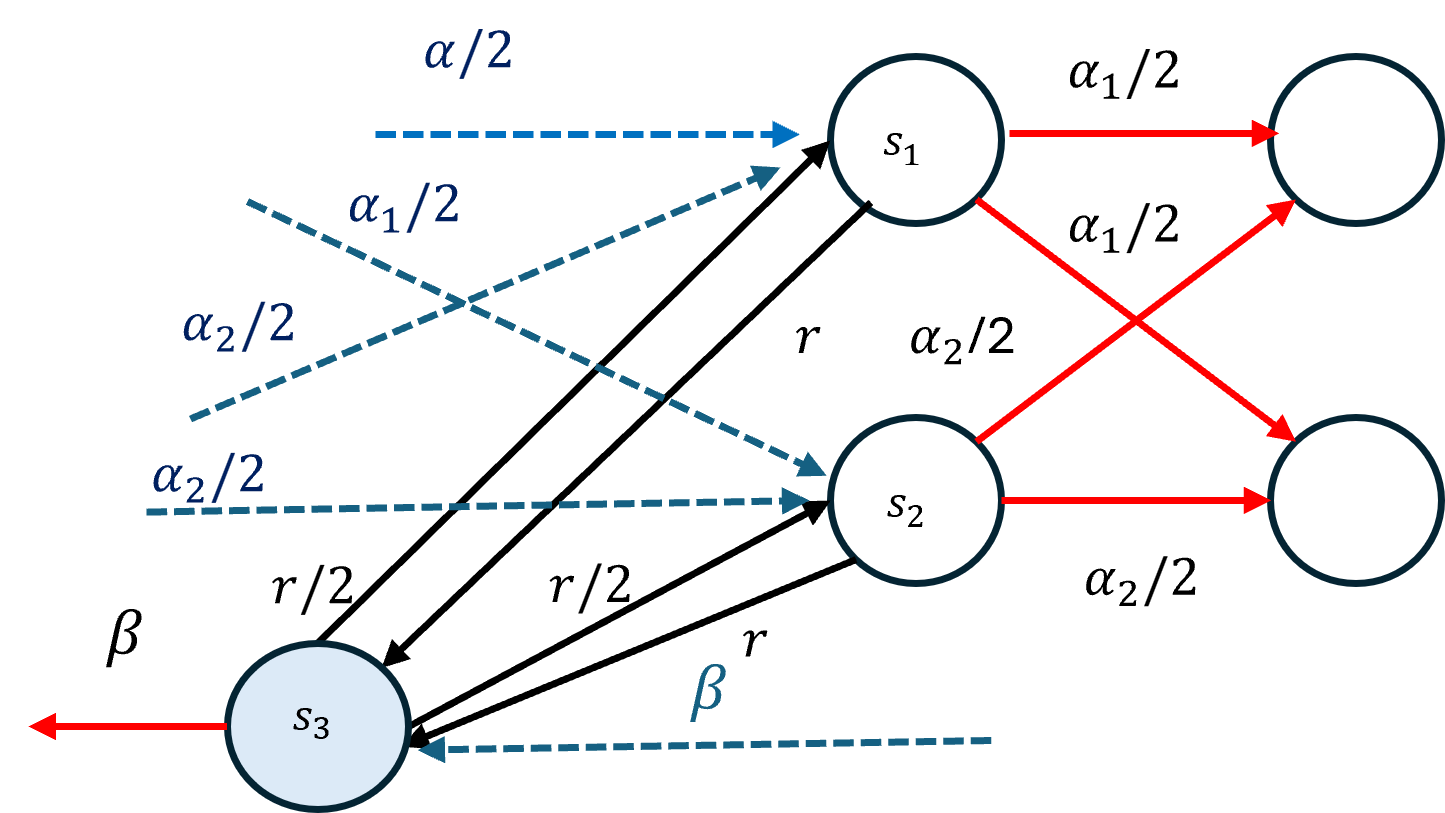}
\caption{}
\label{hyper_ratchet_hmm}
\end{subfigure}
\caption{Hidden Markov cell structure for memory-dependent ratchets. (a) Hypoexponential ratchet. (b) Hyperexponential ratchet.}
\label{ratchet_hmm}
\end{figure}

For the hypoexponential ratchet, a lattice site has an equivalent cell structure with Markov states as shown in figure \ref{ratchet_hmm}(a). The hypoexponential waiting-time distribution of the forward phase of the ratchet is represented by the Markov transitions across the states $s_1$ and $s_2$ connected in series, occurring with constant rates $\alpha_1$ and $\alpha_2$. A transition to the state $s_3$ is possible at each of the sites with reorientation rate $r$. The backwards phase is Markovian, driven only by  exponential waiting-time distributions, therefore no hidden state description is needed; the particle simply transitions out of $s_3$ either to the next state in the backward process with  rate $\beta$ or to $s_1$ in the forward phase with rate $r$. The cell structure in figure \ref{ratchet_hmm}(a) is repeated to generate the equivalent hidden-Markov lattice for the hypoexponential ratchet with periodic boundaries. In the three-site example shown in figure \ref{ratchet_rep} in the main text, there are three cells each containing two states in the outer ring and one in the inner ring. Thus, the transition dynamics can be defined by a $9 \times 9$ generator matrix; via exponential weighting, we finally obtain the tilted generator

\footnotesize
\begin{align}
\label{tilted_hypo}
&\tilde{G}(s)=  \nonumber \\ & \begin{bmatrix} -\alpha_{1}-r & \alpha_{1} & 0 & 0 & 0 & 0 & r & 0 & 0 \\  0 & -\alpha_{2}-r & \alpha_{2}e^s & 0 & 0 & 0& r & 0 & 0 \\ 0 & 0 & -\alpha_{1}-r & \alpha_{1} & 0 & 0 & 0 & r & 0 \\  0 & 0 & 0 & -\alpha_{2}-r & \alpha_{2} e^{s} & 0 & 0 & r & 0 \\ 0 & 0 & 0 & 0 & -\alpha_{1}-r & \alpha_{1} & 0 & 0 & r \\ \alpha_{2} & 0 & 0 & 0 & 0 & -\beta-r & 0 & 0 & r \\ r & 0 & 0 & 0 & 0 & 0 & -\beta-r & 0 & \beta e^{-s} \\ 0 & 0 & r & 0 & 0 & 0 & \beta e^{-s} & -\beta-r & 0 \\ 0 & 0 & 0 & 0 & r & 0 & 0 & \beta e^{-s} & -\beta-r \end{bmatrix}.
\end{align}
\normalsize
The hidden Markov cell structure for the hyperexponential ratchet is shown in figure \ref{ratchet_hmm}(b). The forward phase now contains hidden states $s_1$ and $s_2$ connected in parallel over equiprobable branches. At each hidden state in the forward phase, the particle may choose to transition to $s_3$ in the backward phase, which again has purely Markovian transitions as in the case of the hypoexponential ratchet. After appropriately weighting the transitions in a three-site ring, we arrive at the tilted generator matrix

\footnotesize

\begin{align}
\label{tilted_hyper}
\tilde{G}&(s) = \nonumber\\ &
\begin{bmatrix} -\alpha_{1}-r & 0 & \frac{\alpha_1 e^s}{2} & \frac{\alpha_1 e^s}{2} & 0 & 0 & r & 0 & 0 \\  0 & -\alpha_{2}-r& \frac{\alpha_2  e^s}{2}& \frac{\alpha_2 e^{s}}{2} & 0 & 0& r & 0 & 0 \\ 0 & 0 & -\alpha_{1}-r & 0 & \frac{\alpha_{1}  e^s}{2} & \frac{\alpha_1 e^s}{2} &0 & r & 0 \\  0 & 0 & 0 & -\alpha_{2}-r & \frac{\alpha_{2} e^{s} }{2} & \frac{\alpha_2 e^{s} }{2} & 0 & r & 0 \\ \frac{\alpha_1 e^s}{2} & \frac{\alpha_1 e^{s}}{2} & 0 & 0 & -\alpha_{1}-r & 0 & 0 & 0 & r \\ \frac{\alpha_{2} e^s}{2} & \frac{\alpha_2 e^s}{2}  & 0 & 0 & 0 & -\alpha_{2}-r & 0 & 0 & r \\ r & 0 & 0 & 0 & 0 & 0 & -\beta-r & 0 & \beta e^{-s} \\ 0 & 0 & r & 0 & 0 & 0 & \beta e^{-s} & -\beta-r & 0 \\ 0 & 0 & 0 & 0 & r & 0 & 0 & \beta e^{-s}& -\beta-r \end{bmatrix}.
\end{align}

\normalsize

We obtain the SCGFs of the memory-induced ratchets by numerically calculating the principal eigenvalues of the tilted generators.

\subsection{Gamma TASEP}
\begin{figure}
\centering
\includegraphics[scale=0.5]{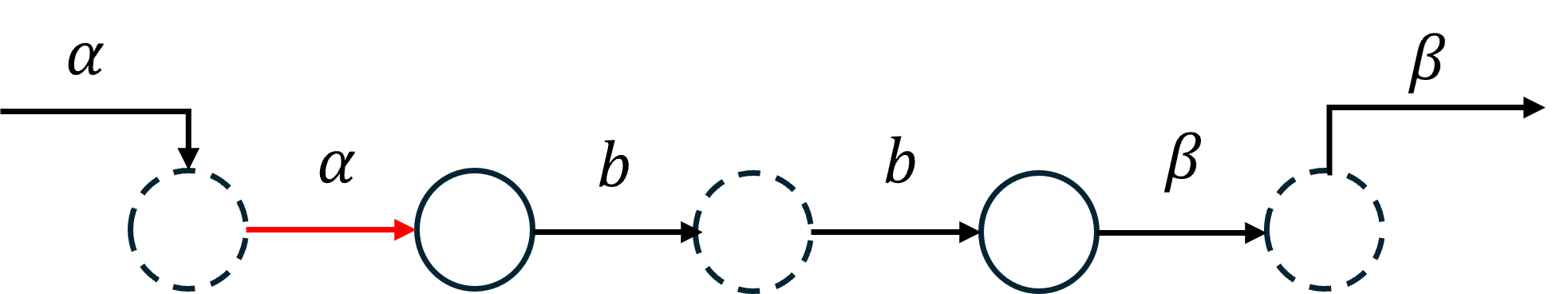}
\caption{Equivalent hidden Markov model for the two-site gamma TASEP. The dotted circles represent the additional hidden sites, giving the fine-grained Markov representation of the coarse-grained gamma-distributed waiting times; $\gamma_{2,\alpha}(\tau)$ for arrivals, $\gamma_{2,b}(\tau)$ for bulk transitions and $\gamma_{2,\beta}(\tau)$ for the departures. The current fluctuations are measured across the red input arrow.}
\label{hmm_tasep}
\end{figure}

In the two-site gamma TASEP of section 5.2.1, the scale parameter, $a=2$, was chosen for all gamma distributed waiting times. In the equivalent Markov representation, a hidden site is added (in series) for each gamma transition in the original lattice. Thus, the two-site gamma TASEP in section 5.2.1 is realized by the five-site hidden process shown in figure \ref{hmm_tasep}.  

The dynamics must now satisfy a two-site exclusion condition, where a particle transition can occur only if two successive sites are unoccupied. This gives ten allowed configurations on the hidden lattice, with a $10 \times 10$ Markov generator in the configuration space.  Appropriately weighting this generator to count current into the system, gives the tilted matrix 

\begin{equation}
\label{tilted_tasep_2}
\tilde{G}(s) = \begin{bmatrix} -\alpha & \alpha & 0 & 0 & 0 & 0 & 0 & 0 & 0 & 0 \\  0 & -\alpha & \alpha e^{s} & 0 & 0 & 0& 0 & 0 & 0 & 0\\ 0 & 0 & -b & b & 0 & 0 & 0 & 0 & 0 & 0\\  0 & 0 & 0 & -b & b  & 0 & 0 & 0 & 0 & 0 \\ 0 & 0 & 0 & 0 & -\alpha-\beta & \alpha & \beta & 0 & 0 & 0 \\ 0 & 0 & 0 & 0 & 0 & -\alpha-\beta & 0 & \alpha e^{s} & \beta  &0\\ \beta & 0 & 0 & 0 & 0 & 0 & -\alpha-\beta & \alpha & 0 & 0 \\ 0 & \beta & 0 & 0 & 0 & 0 & 0& -\alpha-\beta & 0 & \alpha e^{s}\\ 0 & 0 & 0 & 0 & 0 & 0 & 0 & 0  & -\beta & \beta \\ \beta & 0 & 0 & 0 & 0 & 0 & 0 & 0  & 0 & -\beta \end{bmatrix}.
\end{equation}

As in the previous examples, the SCGF is given by the principal eigenvalue of the tilted generator matrix.

\subsection{Form of the alternate dynamics}
It is clear from the construction of the tilted generator matrices in the examples above, that the modified dynamics is still Markovian on the extended (hidden) state space but with different rates to the original process. Thus, the modified dynamics is still semi-Markov on the reduced (visible) state space but with different waiting-time distributions to the original process.
For example, in cases where the original process has a gamma-distributed waiting time, the alternate dynamics (when the conjugate variable $s$ is nonzero) has a hypoexponential waiting-time density. For semi-Markov models without a hidden-variable representation, we believe the alternate process can still be approximated by learning semi-Markov dynamics, albeit with possibly complicated waiting-time distributions. This again justifies the structure used for the two-policy neural actor-critic reinforcement learning.

\section{Robustness and hyperparameter choice}
\label{robustness}
Here we discuss the robustness of the neural reinforcement learning method with respect to important hyperparameters. We choose again the semi-Markov random walk example of section \ref{semi-Markov_example} in the main text for our investigation, anticipating that the picture will be similar for other models. To quantify robustness we study the computational error, defined  as the absolute difference between the SCGF computed via the reinforcement learning algorithm and the corresponding analytically-calculated value. 

\begin{figure}
\centering
\includegraphics[scale=0.5]{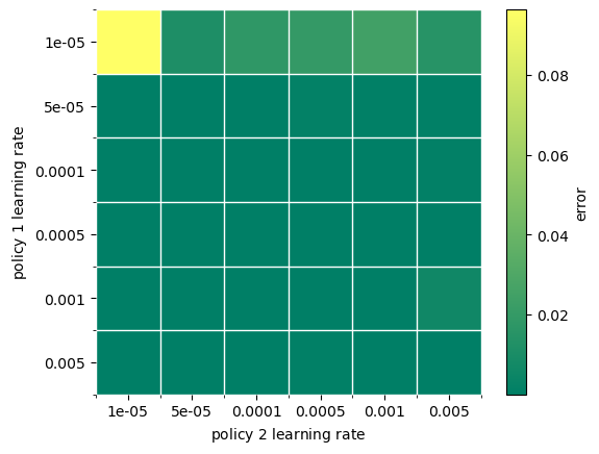}
\caption{Sensitivity of SCGF computations with respect to policy learning rates for the semi-Markov random walk having gamma-distributed waiting times for the forward and backward jumps with parameters  $(a,\beta^{+})=(2,0.6)$ and $(a,\beta^{-}) =(2,0.4)$ respectively. The grid shows the computational error at different learning rates for the two neural actors corresponding to the two policies; policy 1 controls the jumps while policy 2 controls the waiting times. The conjugate variable is here fixed at $s=2$ with  simulation time $t=30\, 000$, while the critic learning rate and the average-reward learning rate are $10^{-4}$ and $ 5 \times 10^{-3}$ respectively.}
\label{learning_policies}
\end{figure}

Figure \ref{learning_policies} shows the variation in the computational error for the semi-Markov random walk with respect to the learning rates for the actors in the two-policy actor-critic, denoted $\alpha_{\theta^p}$ and $\alpha_{\theta^{q}}$ in algorithm \ref{rl_algo}.
To inspect robustness against choice of policy learning rates, the analysis was performed at fixed values for the semi-Markov model parameters, the conjugate variable, and the simulation time. The critic learning rate $\alpha_{\phi}$ and the average-reward learning rate $\alpha_{\bar{r}}$ were also fixed. The grid in figure \ref{learning_policies} shows variation in the errors with the two policy learning rates. A high error is observed when both learning rates are below $10^{-4}$, indicating that effective exploration of the space of network parameters is not possible when the learning rate is too slow. For other values of learning rates shown in the grid, the reinforcement method works efficiently with low errors. However choosing even higher values of learning rates (not shown here) results in learning instabilities and explosion of the gradients.

\begin{figure}
\begin{subfigure}{0.5\textwidth}
\centering
\includegraphics[scale=0.45]{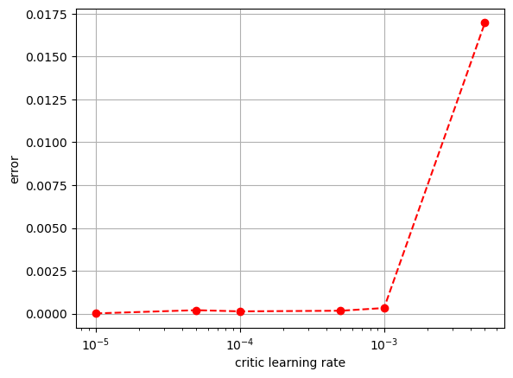}
\caption{}
\label{learning_rate_crtic}
\end{subfigure}
\begin{subfigure}{0.5\textwidth}
\centering
\includegraphics[scale=0.45]{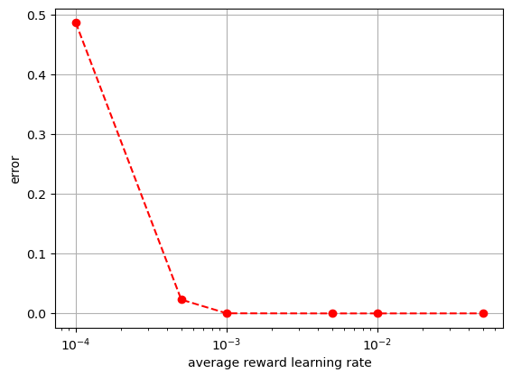}
\caption{}
\label{learning_rate_avg_reward}
\end{subfigure}
\caption{Sensitivity of SCGF computations with respect to the critic and average-reward learning rates for the semi-Markov random walk. (a) Computational error against critic learning rate with average-reward learning rate $5 \times 10^3$. (b) Computational error against average-reward learning rate with critic learning rate $10^{-4}$. Both policy learning rates are  $10^{-4}$, with all other parameters as in figure \ref{learning_policies}. [Dotted lines between data points are a guide to the eye.]}
\label{OAT_sensitivity}
\end{figure}

In figure \ref{OAT_sensitivity}, we show the sensitivity of the learning with respect to the critic and the average-reward learning rates (analyzed one at a time).
The critic and average-reward system have a very similar design to previous implementations for Markovian processes in \cite{rose,diffusive} and sensible choices for the corresponding learning parameters appear similar. Specifically, from figure \ref{OAT_sensitivity}(a), we find an increase in computational error if the critic learning rate is much larger than the learning rates of the two-policies. However, if a much lower critic learning rate than the actors is chosen (not shown in the plot) the average return will not converge to the optimum value.  
From figure \ref{OAT_sensitivity}(b) we see  a high computational error if the average-reward learning rate is lower than the learning rates for the actor-critic networks. 
Beyond the scope of figure \ref{OAT_sensitivity}, choosing very high learning rates leads to instability in gradient computations. 

\begin{figure}
\begin{subfigure}{0.5\textwidth}
\centering
\includegraphics[scale=0.45]{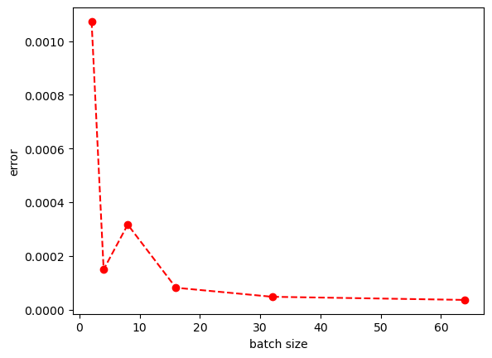}
\caption{}
\label{params_batch_size}
\end{subfigure}
\begin{subfigure}{0.5\textwidth}
\centering
\includegraphics[scale=0.45]{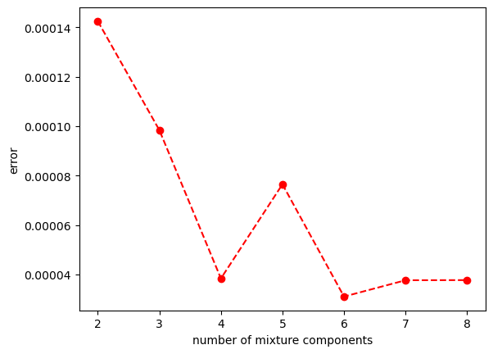}
\caption{}
\label{params_mixture}
\end{subfigure}
\caption{Sensitivity of the reinforcement learning with respect to the batch size and number of gamma-mixture components for the semi-Markov random walk. (a) Computational error against the batch-size with three gamma-mixture components. (b) Computational error against the number of gamma-mixture components with batch-size $16$. Learning rates for both policies and the critic are  $10^{-4}$. The average-reward learning rate is $5 \times 10^{-3}$. All other parameters are as in figure 7. [Dotted lines between data points are a guide to the eye.]}
\label{other_hyperparams}
\end{figure}

For completeness, in figure \ref{other_hyperparams}, we provide plots showing the robustness of the reinforcement learning algorithm against two other hyperparameters: batch size and the number of gamma-mixture components in the actor learning the policy for waiting-time distributions.
As expected, high batch sizes lead to better learning, see figure \ref{other_hyperparams}(a), but are computationally expensive. Similarly, increasing the number of mixture components also tends to reduce the error, see figure \ref{other_hyperparams}(b), but involves a computational cost. Furthermore, increasing the number of mixture components beyond the range shown in figure \ref{other_hyperparams}(b) leads to a high variance in the network outputs and instability in the learning.

The above sensitivity analysis rationalizes the choice of hyperparameters for our implementations. Of course, more complex neural architectures than the one used for the semi-Markov random walk may be more sensitive to the hyperparameters, but the results shown here serve as a preliminary guide. We list the specifics of the hyperparameters chosen with further implementation details for each model in appendix \ref{implementation}.

\section{Recurrent  neural networks for large systems}
\label{RNN method}
For systems with large state spaces, such as the TASEP model of section 5.2.2, the classic feed-forward (dense)  neural networks of section 4.5 are replaced with recurrent  neural networks (RNNs).

\subsection{RNN Structure}

The feed-forward neural network is a weighted connected graph, where each output of the previous layer is connected to the input of the current layer. Adding more layers to the network, or in other words increasing the network \textit{depth}, increases the number of weighted connections, leading to performance drops due to difficulties in gradient computations. Another issue is that feed-forward  neural networks, at the outset, do not include dependencies in the input sequence. We describe next how RNNs address these issues; they are especially useful in problems with sequential data and present a viable framework for processing systems with large state spaces.

\begin{figure}
\centering
\includegraphics[scale=0.5]{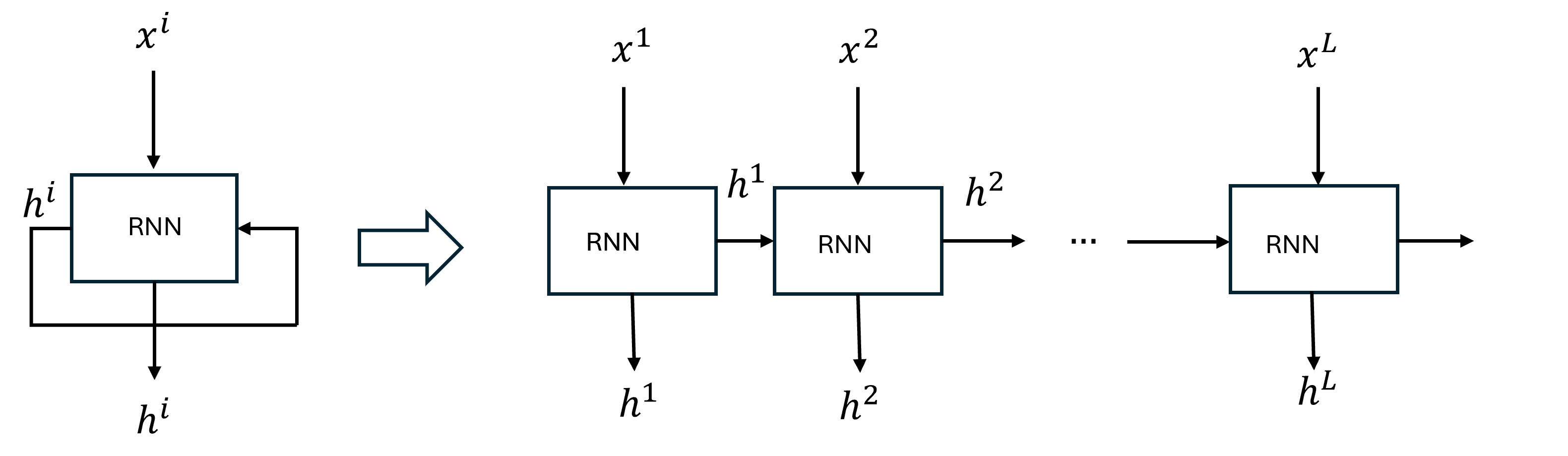}
\caption{Recurrent neural unit. Left-hand side shows a recurrent cell with the corresponding unfolded network shown on the right-hand side.  }
\label{rnn_structure}
\end{figure}

The recurrent neural network, as the name suggests, has a looped structure which is shown in figure \ref{rnn_structure}. Each instance or \textit{time step} of the loop is now considered a \textit{layer} of the network. This is visualized by the unfolded representation of the RNN on the right-hand side of figure \ref{rnn_structure}. A large vector of sequential data is fed component-wise to each layer of the network. At each time step, the internal or hidden state of the RNN is  updated according to the data received and then passed on to the next layer.  The RNN thus propagates information across the layers to learn the sequential correlations in the data. The information flow is controlled by the weights of the RNN. Due to the looped structure, the same set of weights are shared among the layers, reducing the number of parameters to learn.

In practice, popular variants of RNNs such as the so-called long short-term memory  (LSTM) \cite{LSTM,LSTM_review} or the gated recurrent unit (GRU) \cite{gru,gru_eval,GRU_app} are favored over the basic RNN structure, as backpropagation causes the gradient to \textit{vanish} when the number of layers in the basic RNN network becomes large.  In section 5.2.2, we choose to use GRU units that include forgetting and that prioritize information from the most recent layers to update the current internal (hidden) state.  Further details on the machinery of GRU may be found in \cite{gru_eval}. We focus here on the implementation for processing large systems within the context of the reinforcement learning method presented in the main text.

\subsection{GRUs for processing large systems with memory}

In systems like the TASEP of section \ref{many-site TASEP}, obviously  the state space grows with the increase in the number of sites, suggesting that a large inefficient feed-forward network with many layers is required for the neural reinforcement learning. However, since our TASEP states inherently have a sequential structure in the \textit{spatial} dimension, such a large feed-forward network can be avoided by using  a recursive neural network to increase the processing efficiency. 

\begin{figure}
\centering
\includegraphics[scale=0.75]{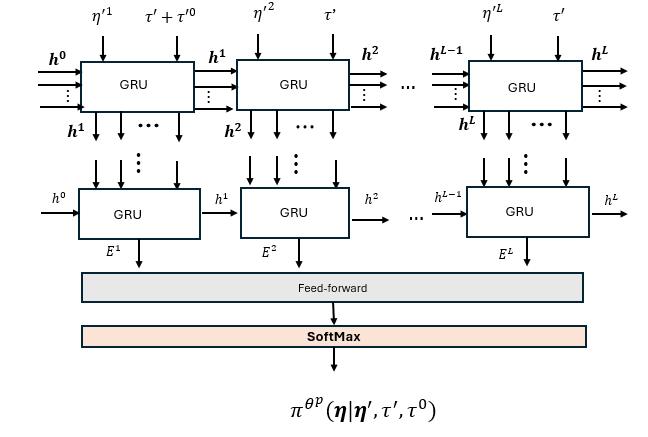}
\caption{A GRU actor architecture for policy $\pi^{p}$. The GRUs systematically process the configuration, waiting time at the present configuration and the time since the arrival clock has been active, to form the encoding $\boldsymbol{E} = [E^1,E^2,...,E^{L}]$. A shallow feed-forward network transforms the encoding to the required transition.}
\label{gru_policy_1}
\end{figure}

\begin{figure}
\centering
\includegraphics[scale=0.75]{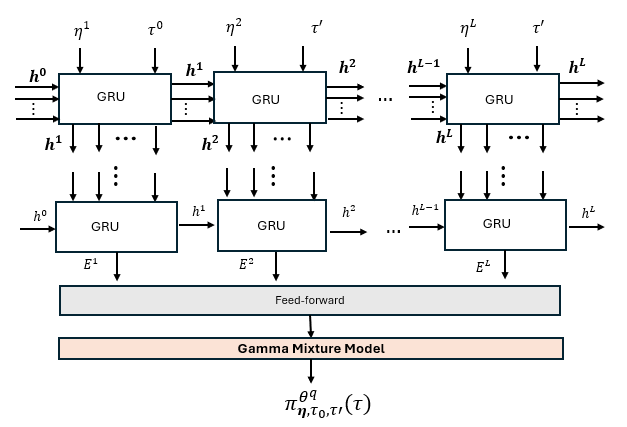}
\caption{A GRU actor architecture for policy  $\pi^{q}$. The GRUs systematically process the present configuration, the waiting time at the previous configuration and the time since the arrival clock has been active to give an encoding $\boldsymbol{E} = [E^1,E^2,...,E^{L}]$. The policy strictly does not need the waiting time $\tau'$, but a marginal improvement was observed when it was included. Again, a shallow feed-forward network transforms the encoding to the required transition.}
\label{gru_policy_2}
\end{figure}

We use GRU-based networks to process the spatial sequence describing the present system state. The actor networks in figures \ref{gru_policy_1} and \ref{gru_policy_2} show how the recurrent architecture can be used to distill the combined information about the configuration and memory into transition policies. The figures show multiple GRUs (unfolded) in a \textit{stacked} architecture, where the outputs of one GRU are fed as inputs to another GRU. For the many-site TASEP, two stacked GRUs were used. The GRU layers transform the information about the current state and memory to create an \textit{encoded} (or \textit{embedded}) vector, $ \mathbf{E}=[E^{1}, E^{2}...,E^{L}]$. The encoding $\mathbf{E}$ can now be used to generate polices with a \textit{shallow} feed-forward network. A similar structure was used for the critic learning the value functions.

\section{Implementation details}
\label{implementation}

The neural networks for all the examples were implemented using pytorch \cite{Pytorch}, a machine learning library for python. Learning rates $(\alpha_{_{\theta^{p}}}, \alpha_{\theta^{q}}, \alpha_{\theta^{v}}) = (10^{-4},10^{-4},10^{-4})$ were chosen for the gradient update of the neural network weights in differential actor-critic of Algorithm~\ref{rl_algo}. The learning rate to update the average reward was $\alpha_{r}=5 \times 10^{-3}$.

As mentioned in the main text, the neural networks were trained on mini-batches. For the semi-Markov CTRW, memory-induced ratchets and two-site gamma TASEP examples, a batch size of $16$ was used. For the actors and the critic in these examples, we used feed-forward neural networks a maximum of three layers deep with $tanh$ activations, followed by a linear layer (without activation). The outputs of these networks were fed into layers performing the specialized operations explained in section \ref{Neural architectures}. We found that this architecture  was sufficient to accomplish the specific learning tasks. For the more complex many-site TASEP, two-stacked GRUs were used for the actors and the critic, followed by a single feed-forward layer with $tanh$ activations and a final linear layer with no activation. For the policies, specialized operations were performed to the outputs of this architecture to obtain the relevant probability distributions. Again, a batch size of $16$ was chosen for all lattices. The networks in this case were trained on a graphics processing unit.

The exclusion conditions in the TASEP models were implemented by masking the relevant outputs (corresponding to the restricted transitions) in the actor learning the policy $\pi^{\theta^{p}}$, as was done in $\cite{gillman2022reinforcement}$, before computing the softmax. The masking ensures the softmax gives zero probabilities for the restricted transitions. The  approximate number of learning parameters in each example is summarized in table \ref{model_params}.

\begin{table}

\begin{tabular}{|c|c|}

\hline
\textbf{Model} & \textbf{No.\ of parameters (approx.)}\\
\hline
Semi-Markov random walk&  $940$ \\
Hypoexponential-exponential ratchet &  $5\,660$  \\

Hyperexponential-exponential ratchet & $5\,660$    \\

2-site TASEP with gamma waiting-times & $7\,821$ \\

2-site TASEP with gamma-distributed arrivals& $8\,904$   \\

10-site TASEP with gamma-distributed arrivals&  $13 \, 008$   \\

64-site TASEP with gamma-distributed arrivals& $40\,710$  \\
\hline
\end{tabular}
\caption{Approximate number of parameters for neural actor-critic models for each example in the main text. The number shown in each case contains the sum of all tunable parameters in the two actors and the critic.}
\label{model_params}
\end{table}


Note that keeping the general overview of the architecture explained in mind, one is free to choose the specific machine learning model parameters, such as number of layers, batch sizes and learning rates. There is much scope for optimization in terms of speed and accuracy, depending on the given nonequilibrium problem and the available hardware. The code for all models is available in a Github repository \cite{Gitcode}.

\section*{Acknowledgments}
The authors would like to thank Dominic Rose (Nottingham) and Massimo Cavallaro (Leicester),  for insightful discussions. RJH benefited from the kind hospitality of the Stellenbosch Institute for Advanced Study (STIAS) during part of this work; she is also grateful for support from the London Mathematical Laboratory in the form of an External Fellowship.

\bibliographystyle{iopart-num}
\bibliography{references}

\end{document}